\documentclass[11pt]{article}
\pdfoutput=1
\usepackage{jheppub}
\usepackage{graphics}
\usepackage{graphicx}
\usepackage{pstricks}
\usepackage{multirow}
\usepackage{slashed}

\DeclareMathOperator{\sgn}{sgn}

\newcommand{\be}{\begin{eqnarray}}
\newcommand{\beq}{\begin{equation}}
\newcommand{\eeq}{\end{equation}}
\newcommand{\ee}{\end{eqnarray}}

\newcommand{\bmp}{\noindent\begin{minipage}{16cm}}
\newcommand{\emp}{\end{minipage}\vskip 7mm} 

\newcommand{\lsim} {\buildrel < \over {_\sim}}
\newcommand{\gsim} {\buildrel > \over {_\sim}}
\usepackage{bm}
\usepackage{amsmath}
\usepackage{amsfonts}
\usepackage{bbm}

\begin{document}
	
\hfill{\tt IFIC/18-12}

\title{Probing secret interactions of eV-scale sterile neutrinos with the diffuse supernova neutrino background}

\author[a]{Yu Seon Jeong,}
\author[b]{Sergio Palomares-Ruiz}
\author[c]{Mary Hall Reno,}
\author[a,d]{Ina Sarcevic,}

\affiliation[a]{Department of Physics, University of Arizona, 1118 E.\ 4th St.\ Tucson, AZ 85704, U.S.A.}
\affiliation[b]{Instituto de F\'{\i}sica Corpuscular (IFIC), CSIC-Universitat de Val\`encia, Apartado de Correos 22085, E-46071 Val\`encia, Spain}
\affiliation[c]{Department of Physics and Astronomy, University of Iowa, Iowa City, Iowa 52242,U.S.A.}
\affiliation[d]{Department of Astronomy, University  of Arizona, 933 N.\ Cherry Ave.,Tucson, AZ 85721,U.S.A.}

\emailAdd{ysjeong@email.arizona.edu}
\emailAdd{sergiopr@ific.uv.es}
\emailAdd{mary-hall-reno@uiowa.edu}
\emailAdd{ina@physics.arizona.edu}

\date{\today}
\abstract{
Sterile neutrinos with mass in the eV-scale and large mixings of order $\theta_0\simeq  0.1$ could explain some anomalies found in short-baseline neutrino oscillation data. Here, we revisit a neutrino portal scenario in which eV-scale sterile neutrinos have self-interactions via a new gauge vector boson $\phi$. Their production in the early Universe via mixing with active neutrinos can be suppressed by the induced effective potential in the sterile sector. We study how different cosmological observations can constrain this model, in terms of the mass of the new gauge boson, $M_\phi$, and its coupling to sterile neutrinos, $g_s$. Then, we explore how to probe part of the allowed parameter space of this particular model with future observations of the diffuse supernova neutrino background by the Hyper-Kamiokande and DUNE detectors. For $M_\phi \sim 5-10$~keV and $g_s \sim 10^{-4}-10^{-2}$, as allowed by cosmological constraints, we find that interactions of diffuse supernova neutrinos with relic sterile neutrinos on their way to the Earth would result in significant dips in the neutrino spectrum which would produce unique features in the event spectra observed in these detectors.
}

\maketitle
\flushbottom

\section{Introduction}

\label{sec:intro}

The past two decades of neutrino oscillation data have proven to be a rich source of information about the masses and mixing in the leptonic sector~\cite{Patrignani:2016xqp}. The standard model three-flavor framework has left-handed neutrinos with mass-squared differences much less than 1~eV$^2$. However, short baseline oscillation experiments like LSND~\cite{Athanassopoulos:1995iw, Athanassopoulos:1996jb, Athanassopoulos:1996wc, Aguilar:2001ty} and MiniBooNE~\cite{AguilarArevalo:2007it, AguilarArevalo:2010wv, Aguilar-Arevalo:2013pmq} and some reactor neutrino experiments~\cite{Mention:2011rk} show anomalies from $\nu_\mu \to \nu_e  \, (\bar\nu_\mu \to \bar\nu_e)$ and $\bar\nu_e$ disappearance, respectively. These anomalous results can be interpreted as pointing to the existence of another species of (sterile) neutrinos with ${\cal O}(1~{\rm eV})$ mass and a mixing angle with an active neutrino species of $\theta_0 \sim 0.1$~\cite{Kopp:2013vaa, Giunti:2013aea}. Indeed, in a framework with three sub-eV active plus one sterile neutrino, global analyses including data from disappearance and appearance experiments obtain as best fit values $\Delta m_{41}^2 = 1.7 $~eV$^2$ and $| U_{e4}|^2 \simeq 0.01-0.02$, $|U_{\mu 4}|^2 \simeq 0.015$ and $|U_{\tau 4}|^2 = 0$~\cite{Collin:2016aqd, Gariazzo:2017fdh, Dentler:2017tkw, Gariazzo:2018mwd}.

With such large mixing angles with active neutrinos, sterile neutrinos would fully thermalize in the bath of Standard Model (SM) particles~\cite{Barbieri:1989ti, Kainulainen:1990ds, Barbieri:1990vx, Enqvist:1991qj, Shi:1993hm, Kirilova:1997sv, Abazajian:2001nj, DiBari:2001ua, Dolgov:2003sg, Melchiorri:2008gq, Hannestad:2012ky, Mirizzi:2012we, Mirizzi:2013gnd, Saviano:2013ktj, Jacques:2013xr, Hannestad:2015tea, Bridle:2016isd} accounting for one extra neutrino at the time of big bang nucleosynthesis (BBN) and the cosmic microwave background (CMB), in tension with current data~\cite{Patrignani:2016xqp}. For $T_\gamma \gg m_4 \equiv m_s$ ($m_4$ is the mass of the mostly sterile neutrino), sterile neutrinos would be relativistic, and their contribution to the relativistic energy density can be parameterized by the effective number of neutrinos, $N_{\rm eff}$. For instance, different cosmological observations, including BBN data, constrain the effective number of fully thermalized neutrinos to be $N_{\rm eff} \lesssim 3.5$ and the mass of the (mostly) sterile neutrinos to be $m_s \lesssim 0.4$~eV (the exact limits depending on the particular data set considered)~\cite{Giusarma:2014zza} (see also, e.g., Refs.~\cite{Hamann:2010bk, Hamann:2011ge, Mirizzi:2013gnd, DiValentino:2013qma, Bergstrom:2014fqa}). This bounds can be satisfied if sterile neutrinos are not in equilibrium or are only partially thermalized, so their contribution to $N_{\rm eff}$ at the relevant epochs is small. 

A possible solution for this problem was suggested long ago in the context of Majoron models~\cite{Babu:1991at, Cline:1991zb} and has been recently revised in scenarios with sterile neutrino self interactions, mediated by a boson~\cite{Hannestad:2013ana, Dasgupta:2013zpn}, sometimes called secret interactions. The new interaction term in the sterile sector would induce an effective potential, which could suppress the large mixing angle in vacuum with active neutrinos and hence, prevent equilibration of sterile neutrinos, and avoiding the bounds from the effective number of relativistic degrees of freedom at BBN. Thus, the production of massive (mostly) sterile neutrinos is suppressed at the epoch of BBN. Even if they recouple with, and again decouple from, active neutrinos before the recombination time, they would satisfy current bounds from both BBN and CMB data~\cite{Hannestad:2013ana, Dasgupta:2013zpn, Saviano:2014esa, Mirizzi:2014ama, Cherry:2014xra, Tang:2014yla, Chu:2015ipa, Archidiacono:2015oma, Cherry:2016jol, Archidiacono:2016kkh}. In recent years, more detailed analyses of cosmological data have been performed and different phenomenological consequences of the idea have been studied~\cite{Bringmann:2013vra, Ko:2014nha, Archidiacono:2014nda, Kopp:2014fha, Saviano:2014esa, Mirizzi:2014ama, Cherry:2014xra, Tang:2014yla, Forastieri:2015paa, Chu:2015ipa, Archidiacono:2015oma, Cherry:2016jol, Archidiacono:2016kkh}. The conclusion is that as a result of these new interactions, BBN and CMB limits permit a relic density of sterile neutrinos with a mass of about 1~eV.

On the other hand, the coupling of neutrinos to a mediator could give rise to an attenuated spectrum of cosmic neutrinos due to the resonance production of this mediator in the relic neutrino background~\cite{PalomaresWeiler, Weilertalk, PalomaresRuiztalk07, PalomaresRuiztalk11, Hooper:2007jr, Ioka:2014kca, Ng:2014pca}, an idea similar to the attenuation of the flux of ultra-high energy neutrinos due to the resonant interaction with cosmic relic neutrinos at the $Z$-pole~\cite{Weiler:1982qy, Weiler:1983xx, Roulet:1992pz, Yoshida:1996ie, Eberle:2004ua, Barenboim:2004di}. Similarly, the diffuse supernova neutrino background (DSNB) flux could also experience distortions en route to Earth, either because of interactions with the relic neutrino background in models with additional $Z'$ gauge bosons coupled to neutrinos~\cite{Goldberg:2005yw, Baker:2006gm} or because of interactions with dark matter particles~\cite{Farzan:2014gza} in models with radiatively-generated neutrino masses~\cite{Boehm:2006mi, Farzan:2009ji, Farzan:2010mr, Farzan:2011tz, Farzan:2012ev}.

In the scenario discussed in this paper (i.e., an extra eV-scale sterile neutrino with self interactions mediated by a vector boson $\phi$), the relic density of sterile neutrinos could act as the target for the DSNB flux, which would resonantly produce the vector boson $\phi$. Thus, we would also expect a dip in the event spectrum from the DSNB if the resonant energy of this interaction lies in the relevant range for supernova (SN) neutrinos (i.e., tens of MeV). For targets with eV mass, absorption features could show up for mediators with masses in the keV range. Moreover, note that for masses in the MeV range, these secret interactions in the sterile neutrino sector could also produce dips in the cosmic neutrino spectrum~\cite{Cherry:2014xra, Cherry:2016jol}.

Here we focus on the expected signals by the DSNB in the liquid argon (LAr) detector planned for the Deep Underground Neutrino Experiment (DUNE)~\cite{Acciarri:2015uup, Acciarri:2016crz} and in the water-\v{C}erenkov Hyper-Kamiokande (HK) detector~\cite{HK}, in scenarios with self interactions of sterile neutrinos with eV masses and relatively large vacuum mixing with active neutrinos ($\theta_0 \sim 0.1$).

The structure of the paper is as follows. In Section~\ref{sec:lagrangian}, we describe the low-energy Lagrangian in the sterile neutrino sector and its $V-A$ interactions with a new vector boson. We introduce the sterile neutrino production rate in the early Universe and describe its ingredients in detail: the collision rate and the average probability for active-sterile neutrino conversions in the medium, which depend on the effective potential induced by these new interactions and on the quantum damping rate. We provide a compendium of the relevant cross sections, including thermal averaging, relevant to the cosmological constraints reviewed and discussed in Section~\ref{sec:constraints}. The different models we consider for the DSNB flux are described in Section~\ref{sec:DSNB}, as well as the signals expected in future DUNE and HK detectors with and without self interactions of sterile neutrinos. In Section~\ref{sec:conclusions} we discuss our results and draw our conclusions. Finally, in Appendix~\ref{app:Veff} we include the details of the calculation of the effective potential due to the new interactions in the sterile neutrino sector.

\section {Sterile neutrino interactions}
\label{sec:lagrangian}

Here we consider a scenario with one extra sterile neutrino which has self-interactions mediated by a new keV-MeV scale $U(1)$ vector boson ($\phi$) and has interactions with the SM sector only via mixing with active neutrinos. In this section, we review the interaction rates considered in different detail in the literature~\cite{Hannestad:2013ana, Dasgupta:2013zpn, Cherry:2014xra, Chu:2015ipa, Cherry:2016jol}. We include thermal averaging of the cross sections, study the impact of the resonant scattering $\nu_s \bar{\nu}_s \to \phi \to \nu_s \bar{\nu}_s$ and of the effective potential $V_{\rm eff}$, that enters the sterile-active mixing in the medium. The interaction rates and effective mixing angle are inputs to determine cosmologically allowed regions in the coupling constant--vector boson mass parameter space. 

In this work, cosmologically relevant interactions in the sterile sector are assumed to occur between sterile neutrinos and a vector boson $\phi$, described by the interaction term~\cite{Dasgupta:2013zpn, Chu:2015ipa}
\begin{equation}
\label{eq:Ls}
{\cal L}_{s} = g_s \, \bar{\nu}_s \gamma_\mu  P_L \, \nu_s \, \phi^\mu ~,
\end{equation}
where $P_L = (1 - \gamma_5)/2$. The $V - A$ coupling keeps the number of degrees of freedom for sterile neutrinos plus antineutrinos the same as for active neutrinos, namely, $g_{\nu_s} = g_{\nu_a} = 2$.\footnote{We use $g_s$ to denote the coupling and $g_{\nu_s}$ for the number of degrees of freedom of the sterile neutrino.} We also assume that this sector ($\nu_s,\ \bar{\nu}_s$ and $\phi$) decouples from SM particles above the TeV scale, when the effective number of degrees of freedom is on the order of $g_*\sim 106.7$~\cite{Kolb:1990vq}. If particles in the sterile sector do not recouple to SM particles (e.g., via mixing) before BBN, the early decoupling ensures that the number density of sterile neutrinos does not equal that of active neutrinos (due to the entropy release in the SM sector), so that the constraint on the effective number of extra neutrinos during the BBN epoch, namely that $\Delta N_{\rm eff} $ is a fraction of a SM neutrino species at $T_\gamma \lsim 1$~MeV, is satisfied~\cite{Cyburt:2015mya, Pitrou:2018cgg} (see also Ref.~\cite{Cooke:2017cwo} for a less constraining limit). We use $N_{\rm eff}^{\rm BBN}\lsim 3.2$ during the BBN epoch for our constraints.
\footnote{While there could be potential constraints on active-sterile oscillations after (or around) neutrino decoupling through the distortion of the active neutrino energy distributions~\cite{Hannestad:2013ana, Saviano:2014esa}, a more detailed discussion on this regard is beyond the scope of this paper.}

With these assumptions, neglecting the impact of the active-sterile mixing, the ratio of the sterile neutrino temperature $T_s$ to the active neutrino temperature $T_\nu$, $\xi \equiv {T_s}/{T_\nu}$ at $T_\gamma \sim 1$~MeV depends on whether or not $\phi$'s are present. 

If the $\phi$'s are relativistic at BBN\footnote{Given that we are not solving the Boltzmann equations, the exact value for the transition from relativistic to non-relativistic $\phi$ is set by matching the constraints presented below in the two regimes.} (i.e., $M_\phi \lsim 1$~MeV), they are easily produced with a temperature equal to that of sterile neutrinos. Therefore, the temperature ratio $\xi$ and the effective number of neutrino species during BBN are
\begin{eqnarray}
\label{eq:xiA}
\xi _{\rm rel} & = & \Biggl(\frac{10.75}{106.75}\Biggr)^{1/3} \simeq 0.465 \\
N_{\rm eff}^{\rm rel} & = & N_{\nu_a}+ \frac{g_{\nu_s} \cdot 7/8  + g_\phi}{g_{\nu_a} \cdot 7/8} \, \xi_{\rm rel}^4 \simeq 3.17 ~,
\end{eqnarray}
where we have used the SM value, $N_{\nu_a} = 3.045$~\cite{deSalas:2016ztq} (see also earlier calculations~\cite{Mangano:2001iu, Mangano:2005cc}) and $g_\phi = 3$. 

If $\phi$'s are non-relativistic (i.e., $M_\phi \gsim 1$~MeV), they would have decayed away into sterile neutrinos by the BBN epoch and thus,
\begin{eqnarray}
\label{eq:xiB}
\xi _{\rm nr} & = & \left(\frac{10.75}{106.75}\right)^{1/3}\left(\frac{2\cdot 7/8 + 3}{2 \cdot 7/8}\right)^{1/3} \simeq 0.649 \\
N_{\rm eff}^{\rm nr} & = & N_{\nu_a}+ \xi_{\rm nr}^4\simeq 3.22 ~,
\end{eqnarray}
where the second factor in $\xi_{\rm nr}$ accounts for the `heating' of sterile neutrinos from $\phi$ decays. Note that this factor is not present in Eq.~(\ref{eq:xiA}), where the temperature ratio only corresponds to the SM entropy release between high temperatures ($\gsim 1$~TeV) and BBN temperatures ($\lsim 1$~MeV), with no `heating' in the sterile sector. Therefore, these results represent the two limiting cases for different $M_\phi$. As we can see, regardless the value of $M_\phi$, the BBN constraint $N_{\rm eff}^{\rm BBN} \lsim 3.2$~\cite{Cyburt:2015mya, Pitrou:2018cgg,Cooke:2017cwo} is always satisfied if the sterile sector does not recouple to the SM sector before $T_\gamma \sim 1$~MeV. If recoupling occurs before that time, the bounds from BBN would be violated.  We will use the ratio of temperatures $\xi$ given in Eq.~(\ref{eq:xiA}) or ~(\ref{eq:xiB}), for $M_\phi < 1$~MeV or $M_\phi \geq 1$~MeV, respectively, in our evaluation of cosmological constraints on the sterile sector. Obviously, this is a rough approximation, although the changes in the results are not qualitatively important.

Large mixing between sterile and active neutrinos would drive the former to reach thermal equilibrium in the early Universe, i.e., to recouple and thus, there would be an extra contribution to the number of neutrino degrees of freedom, violating cosmological (BBN and CMB) bounds. Active-sterile mixing in the medium depends on the vacuum mixing angle $\theta_0$ and on the effective potential $V_{\rm eff}$~\cite{Notzold:1987ik} (see below). In the case of no extra new interaction in the sterile sector, the SM weak potential is negligible at MeV temperatures or below, so the mixing angle is the one in vacuum. Nevertheless, if a term like Eq.~(\ref{eq:Ls}) is present, there is an extra contribution to the sterile neutrino self-energy, which can dominate over SM matter effects, even for small values of $g_s$. This is so because the mass of the $\phi$ under consideration here is much smaller than the $W$ mass. 

The production rate of sterile neutrinos, $\Gamma_{\nu_s}$, is given by the product of half the total interaction rate~\cite{Stodolsky:1986dx, Thomson:1991xq} times the thermal average of the active-sterile neutrino conversion probability~\cite{Foot:1996qc}
\begin{equation}
\label{eq:Gs}
\Gamma_{\nu_s} (\nu_a \to \nu_s) = \frac{\Gamma_{\rm int} }{2} \, \langle P (\nu_a \to \nu_s) \rangle ~.
\end{equation}

The average probability for active-sterile neutrino conversions in the medium (in the adiabatic limit) can be written as~\cite{Foot:1996qc, Volkas:2000ei}
\begin{equation}
\label{eq:Pnunu}
\langle P (\nu_a \to \nu_s) \rangle 
\simeq \frac{1}{2} \frac{\frac{\Delta m_s^2}{2 \, E} \, \sin^2 2 \theta_0}{(\frac{\Delta m_s^2}{2 \, E} \, \cos 2 \theta_0 + V_{\rm eff} )^2
	+ \frac{\Delta m_s^2}{2 \, E} \, \sin^2 2 \theta_0 +  D_{\rm int}^2}  ~,
\end{equation}
where the overall factor of $1/2$ is the result of averaging the oscillatory term, $\theta_0$ is the vacuum mixing angle and $\Delta m_s^2$ is the (mostly) active-sterile neutrino mass difference squared. In what follows, we take $\theta_0 = 0.1$ and $\Delta m_s^2 = 1$~eV$^2$.

There are two important terms in Eq.~(\ref{eq:Pnunu}) which are present because oscillations take place in a medium: $V_{\rm eff}$ is the effective potential induced by neutrino forward scattering in the thermal bath (both from SM interactions and from the new sterile sector interactions) and $D_{\rm int} = \Gamma_{\rm int} / 2$ is the quantum damping rate and accounts for the loss of coherence due to collisions~\cite{Stodolsky:1986dx}. 

The effective potential $V_{\rm eff}$ receives finite temperature contributions from both SM and new sector interactions. Due to mixing between active and sterile neutrinos, SM interactions also contribute to the effective potential of (mostly) sterile neutrinos, although suppressed by four powers of the mixing angle, $\sin^4 \theta_0$. Similarly, interactions in the sterile sector, Eq.~(\ref{eq:Ls}), also contribute to the effective potential of active neutrinos. Therefore, the effective potential appearing in Eq.~(\ref{eq:Pnunu}) can be written as
\begin{equation}
V_{\rm eff} = \left[V_{\rm eff, s}(E_s, T_s) - V_{\rm eff, SM} (E_\nu, T_\nu)\right] + \sin^4 \theta_0 \, \left[V_{\rm eff, SM} (E_s, T_\nu)  - V_{\rm eff, s} (E_\nu, T_s) \right] ~.
\end{equation}

The SM contribution to the effective potential at temperatures below the mass of the SM gauge bosons was computed three decades ago and is given by~\cite{Notzold:1987ik}
\begin{eqnarray}
\label{eq:VeffSM}
V_{\rm eff, SM} (E, T_\nu) & \simeq & - \frac{8 \, \sqrt{2}}{3} \, G_F \, E \, \left( 
\frac{\langle E_\nu \rangle \, n_\nu + \langle E_{\bar\nu} \rangle \, n_{\bar\nu}}{m_W^2} 
+ \kappa \, \frac{\langle E_\ell \rangle \, n_\ell + \langle E_{\bar\ell} \rangle \, n_{\bar\ell}}{m_Z^2} \right) \nonumber \\
& \simeq & - 4.3 \, G_F \, \left(\frac{1}{m_W^2} + \frac{\kappa}{m_Z^2}\right) \, E \, T_\nu^4  ~, 
\end{eqnarray}
where, at the temperatures $1~{\rm MeV} \lsim T_\nu \lsim 100$~MeV, $\kappa = 1$ for $\nu_e$ and $\kappa = 0$ for $\nu_\mu$ and $\nu_\tau$ (as there are no $\mu$ or $\tau$ leptons at $T_\nu < 100$~MeV). For numerical computations we use the electron neutrino case, although the differences do not affect our discussion.

Although we use the full expression, Eq.~(\ref{eq:Veff2}) in Appendix~\ref{app:Veff}, the effective potential from interactions in the sterile sector, Eq.~(\ref{eq:Ls}), can be analytically computed in the low- and high-temperature limits~\cite{Dasgupta:2013zpn}, 
\begin{eqnarray}
\label{eq:Vefflim}
V_{\rm eff, s} (E, T_s) \simeq 
\begin{cases}
- \frac{7 \, \pi^2 \, g_s^2}{45} \, \frac{E \, T_s^4}{M_\phi^4} \hspace{1cm} {\rm for} \quad T_s \ll M_\phi ~, \\[1ex]
\,  \frac{g_s^2}{8} \, \frac{ T_s^2}{E} \hspace{2.1cm}  {\rm for} \quad T_s \gg M_\phi ~,
\end{cases}
\end{eqnarray}
where $E$ is the sterile neutrino energy (which has to coincide with that of active neutrinos so that the two states can oscillate) and equal distributions for active neutrinos and antineutrinos are assumed. In the Appendix, we include the expression for $V_{\rm eff, s} (E, T_s)$ corresponding to light sterile neutrinos, which is also applicable near the resonant production of $\phi$. When the effective potential is more important than the vacuum term, the probability for the active-sterile neutrino conversion is suppressed, preventing the equilibration of sterile and active neutrinos and preserving the consistency with the BBN limit on effective number of extra neutrinos~\cite{Hannestad:2013ana, Dasgupta:2013zpn}. 

The total interaction rate, which appears in the definition of the sterile neutrino production rate, $\Gamma_{\nu_s}$, and in the damping rate, $D_{\rm int}$, can be written as
\begin{equation}
\Gamma_{\rm int} \equiv \Gamma_{\rm int, SM} + \Gamma_{\rm int, s} = \langle \sigma_a v_{\rm Mol} \rangle \, n_a + \langle \sigma_s v_{\rm Mol} \rangle \, n_s  + \langle \sigma_\phi v_{\rm Mol} \rangle \, n_\phi ~,
\label{eq:GiNs}
\end{equation}
where $n_a$, $n_s$ and $n_\phi$ are the number densities and $\langle \sigma_a v_{\rm Mol} \rangle$, $\langle \sigma_s v_{\rm Mol} \rangle$ and $\langle \sigma_\phi v_{\rm Mol} \rangle$ are the thermal average of the cross section times the M\o{}ller velocity (equal to the relative velocity in the lab or center-of-mass frames), corresponding to active and sterile neutrinos and $\phi$ bosons, respectively. The total interaction rate of sterile neutrinos, $\Gamma_{\rm int}$, has three contributions: the usual one from SM interactions of active neutrinos, the ones from collisions of sterile neutrinos or between sterile neutrinos and $\phi$ bosons due to the new interaction term. In the first case, with SM interactions, sterile neutrino production proceeds via active-active neutrino interactions and then active-sterile neutrino mixing, so the temperature of the final sterile neutrino state is that of active neutrinos. In the second and third cases, with new sterile neutrino interactions, sterile neutrinos are produced via active-sterile neutrino mixing and then sterile-sterile neutrino and sterile neutrino-$\phi$ boson interactions, respectively, so the temperatures of the distributions of the incoming active and sterile neutrinos are, in principle, different, $T_\nu \neq T_s$. Sterile neutrino-$\phi$ interactions 
do not change our conclusions here, so for simplicity, we do not discuss them further here.

The contribution from active-active neutrino interactions, in the case of active-sterile neutrino oscillations, is given by~\cite{Enqvist:1991qj}
\begin{equation}
\Gamma_{\rm int, SM} = y_\alpha \, G_F^2 \, T_\nu^5 ~,
\label{eq:GintSM}
\end{equation}
where $y_\alpha \simeq 4.1$ for $\alpha = \nu_e$ and $y_\alpha \simeq 2.9$ for $\alpha = \nu_\mu, \nu_\tau$. For numerical computations we consider the electron neutrino case, although very similar results are obtained otherwise.

For $\Gamma_{\rm int,s}$, the contribution to the total interaction rate arising entirely from the sterile neutrino sector, we have to consider both elastic and inelastic interactions of sterile neutrinos induced by the new term in the Lagrangian, Eq.~(\ref{eq:Ls}),
\begin{eqnarray}
\nu_s \, \bar{\nu}_s & \to  &\nu_s \, \bar{\nu}_s  ~, \\
\nu_s \, \nu_s & \to  &\nu_s \, \nu_s  ~, \\
\nu_s \, \bar{\nu}_s & \to & \phi \, \phi ~, 
\end{eqnarray}
and similarly for $\bar{\nu}_s$.

The first process $\nu_s(p_1) \, \bar{\nu}_s (p_2) \to  \nu_s(p_3) \, \bar{\nu}_s(p_4)$ is equivalent to Bhabha scattering of electrons, with the substitution of $V - A$ couplings of a massive boson for the vector couplings of the photon. The cross section has both $s$- and $t$-channel contributions to the matrix element squared, where $s = (p_1 + p_2)^2$ and $t = (p_1 - p_3)^2$ are the Mandelstam variables. Although in our calculations we use the full expression, in various limits, the cross section for $\nu_s \, \bar{\nu}_s  \to \nu_s \, \bar{\nu}_s$ and $\phi$ decay width read
\begin{eqnarray}
\label{eq:sigs}
\sigma_s \equiv \sigma ({\nu_s  \, \bar{\nu}_s  \to \nu_s \, \bar{\nu}_s}) & = &
\begin{cases}
\frac{ g_s^4}{ 4 \, \pi \, M_\phi^2}  \hspace{3cm} {\rm for} \quad \, s > M_\phi^2 ~,  \\
\frac{g_s^4}{12 \, \pi} \, \frac{s}{\left(s - M_\phi^2\right)^2 + M_\phi^2 \, \Gamma_\phi^2} \hspace{0.9cm} {\rm for} \quad  s \sim M_\phi^2  ~, \\
\frac{g_s^4}{3 \, \pi \, M_\phi^4} \, s  \hspace{2.8cm} {\rm for} \quad \, s < M_\phi^2 ~,   \\
\end{cases}
\\[2ex]
\Gamma_\phi & = &\frac{g_s^2 \, M_\phi}{24 \, \pi}~.
\end{eqnarray}
Note that in the low-mass limit, $s > M_\phi^2$, the $t$-channel cross section depends on $M_\phi^{-2}$, instead of $E^{-2}$, as considered in Refs.~\cite{Archidiacono:2014nda, Mirizzi:2014ama, Chu:2015ipa}. This had already been noted in Refs.~\cite{Cherry:2014xra, Cherry:2016jol}.

The thermal average of the cross section times relative velocity $v_{\rm Mol}$ is given by~\cite{Gondolo:1990dk}
\begin{equation}
\langle  \sigma v_{\rm Mol}\rangle  \simeq
\frac{\int \sigma v_{\rm Mol} \, f_1 \, f_2 \, d^3 p_1 \, d^3 p_2 }
{\int f_1 \, f_2 \, d^3 p_1 \, d^3 p_2 }  ~,
\label{eq:ThAvg}
\end{equation}
where $f_{i}(E_i) = \left(e^{(E_i - \mu_i)/T_i} \pm 1\right)^{-1}$ is the distribution function of species $i$. We assume neutrinos and antineutrinos are equally distributed and thus, $\mu_i = 0$. The M\o{}ller velocity, $v_{\rm Mol}$, is defined as 
\begin{equation}
v_{\rm Mol} \equiv \frac{[(p_1 \cdot p_2)^2 - m_1^2 \, m_2^2]^{1/2}}{E_1 \, E_2} ~, 
\end{equation}
which is approximated by $v_{\rm Mol} \simeq 1-\cos \theta$ in the relativistic limit ($p_i \simeq E_i$), valid for most of our discussion since all neutrino masses are of the order of 1~eV or below. Thermal averaging with Eq.~(\ref{eq:ThAvg}) is essential for the proper treatment of the resonance behavior of the annihilation cross section. From the cross sections in Eq.~(\ref{eq:sigs}), the analytic expression of thermal average for the high- and low-energy limits is
\begin{equation}
\label{eq:sig-th}
\langle \sigma_s v_{\rm Mol}\rangle_{\nu_s \bar\nu_s} \simeq 
\begin{cases}
\frac{\pi \, g_s^2 \, M_\phi^2}{18 \, T_1^2 \, T_2^2 \, \zeta (3)^2 } + \frac{g_s^4}{4 \, \pi \, M_\phi^2} 
\hspace{1cm} {\rm for} \quad s > M_\phi^2 ~,   \\[2ex] 
\frac{4 \, g_s^4 \, T_1 \, T_2}{M_\phi^4 \, \zeta(3)^2} \hspace{3.1cm}  {\rm for} \quad s \ll M_\phi^2  ~,
\end{cases}
\end{equation}
where $T_1$ and $T_2$ are the temperatures of the two neutrino distributions, which can be different due to active-neutrino mixing. For $T \lsim M_\phi$, the thermally averaged ($s$-channel) cross section increases rapidly. A simple form in this case is not available. We use the numerical results below.

\begin{figure}[t]
	\centering
	\includegraphics[width=0.9\textwidth]{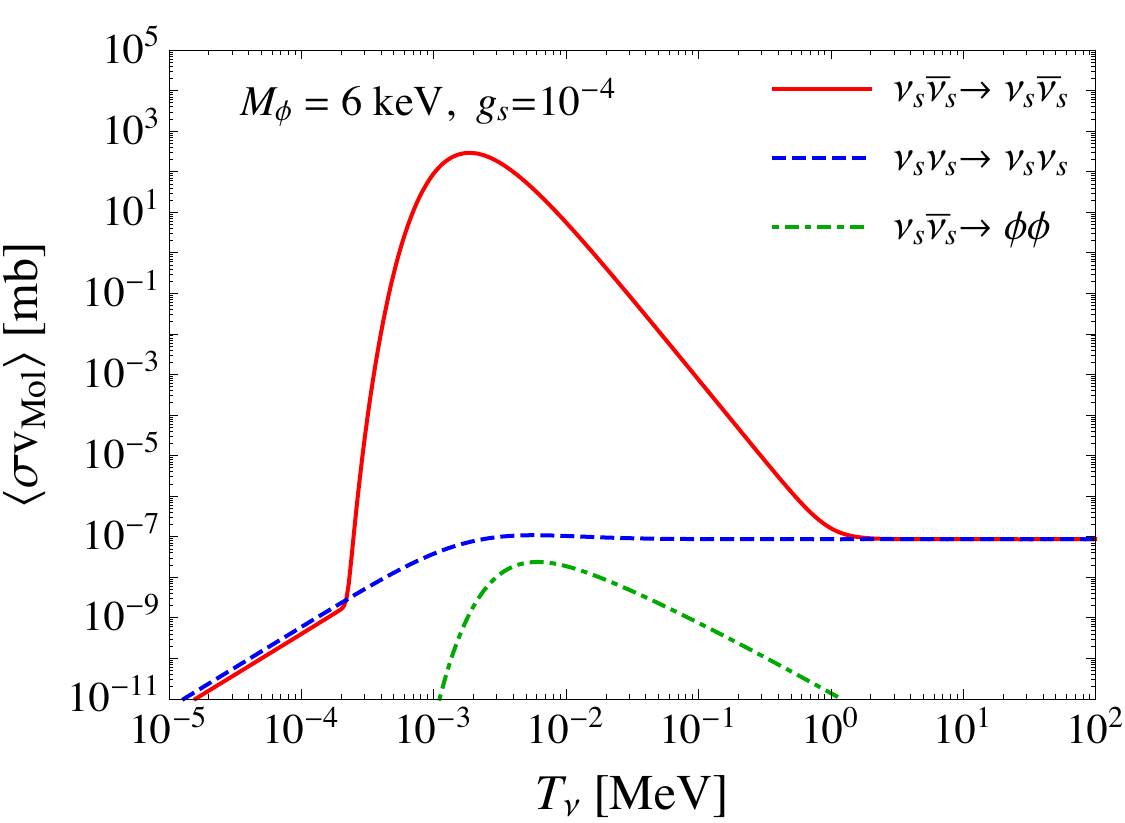}
	\caption{ 
		Thermal average of the sterile neutrino cross sections, $\langle \sigma v_{\rm Mol}\rangle$, as a function of the active neutrino temperature $T_\nu$ with  $T_1 = T_s = 0.465 \, T_\nu$ and $ T_2 = T_\nu$ in Eq.(\ref{eq:sigcross-th}), for $\nu_s \, \bar{\nu}_s \to \nu_s \, \bar{\nu}_s$ (upper red solid curve) and $\nu_s \, \bar{\nu}_s \to \phi \, \phi$ (green dot-dashed curve). The $\nu_s \, {\nu}_s \to \nu_s \, {\nu}_s$ thermally averaged process (blue dashed curve) is also shown. The input parameters are the default values used in Section~\ref{sec:DSNB}: $M_\phi = 6$~keV and $g_s = 10^{-4}$.
	}
	\label{fig:sigs-avg}
\end{figure}

The corresponding result for $\nu_s \, \nu_s \to \nu_s \, \nu_s$ can be obtained by using crossing symmetry. In the high- and low-energy limits, it is given by
\begin{equation}
\label{eq:sigcross-th}
\langle \sigma_s v_{\rm Mol}\rangle_{\nu_s \nu_s} \simeq 
\begin{cases}
\frac{g_s^4}{4 \, \pi \, M_\phi^2} 
\hspace{2.3cm} {\rm for} \quad s > M_\phi^2 ~,   \\[2ex] 
\frac{6 \, g_s^4 \, T_1 \, T_2}{M_\phi^4 \, \zeta(3)^2} \hspace{2cm}  {\rm for} \quad s < M_\phi^2  ~,
\end{cases}
\end{equation}

For the $\nu_s \, \bar{\nu}_s \to \phi \, \phi$ process, the cross section is
\begin{equation}
\label{eq:sstophph}
\sigma (\nu_s \, \bar{\nu}_s \to \phi \, \phi)  =\frac{g_s^4}{2 \, \pi \, s} 
\left[ \left(1+ \frac{4 \, M_\phi^4}{s^2} \right) \left(1- \frac{2 \, M_\phi^2 }{s} \right)^{-1} 
\ln \Biggl[ \frac{1 + \beta}{1-\beta} \Biggr] - \beta  \right] ~,
\end{equation}
with 
$$
\beta = \sqrt{1-\frac{4 \, M_\phi^2}{s}}  \hspace{1cm} {\rm and} \qquad s > 4 \, M_\phi^2 ~.
$$
The $\nu_s \, \phi \to \nu_s \, \phi$ cross section can be obtained in a similar way, although collinear divergences have to be taken care of. In any case, both $\nu_s \, \bar{\nu}_s \to \phi \, \phi$ and $\nu_s \, \phi \to \nu_s \, \phi$ are subdominant processes at all temperatures.

We show in Fig. \ref{fig:sigs-avg} the thermal averaged annihilation and scattering cross sections, as a function of the standard model neutrino temperature, for $g_s = 10^{-4}$ and $M_\phi = 6$~keV. The cross sections are relevant to the cosmological constraints when a sterile neutrino scatters with an active neutrino, via mixing. The thermally averaged total cross section for $\nu_s\bar{\nu}_s \to \nu_s\bar{\nu}_s$ process (upper red solid curve) results from the sum of the $t$-channel and $s$-channel plus interference term. The scattering process $\nu_s\nu_s \to \nu_s\nu_s$ is shown by the blue dashed curve ($t$-channel). We also show the sterile neutrino annihilation into $\phi \phi$ (green dot-dashed curve). As can be seen from the figure, only the cross sections for $\nu_s \, \bar\nu_s \to \nu_s \, \bar\nu_s$ and $\nu_s \, \nu_s \to \nu_s \, \nu_s$ are relevant to determine whether sterile neutrinos are in thermal equilibrium or not.

\section{Cosmological constraints on $(M_\phi, g_s)$}
\label{sec:constraints}

As discussed above, we assume the sterile sector particles are present in the early Universe, but that they decouple from active neutrinos at temperatures well above the electroweak scale. In addition to the primordial population, sterile neutrinos can be produced from interactions with active neutrinos via mixing. A detailed evaluation of the sterile neutrino abundance would require a solution to the quantum kinetic equations for the momentum-dependent density matrix (see, e.g., Refs.~\cite{Kainulainen:2001cb, Hannestad:2015tea}). Nevertheless, we use the ratio of the sterile neutrino production rate to the Hubble expansion rate to establish whether or not sterile neutrinos equilibrate at a given temperature. We make the approximation that when the production rate is higher than the Hubble expansion rate, thermal equilibrium is established between sterile and active neutrinos. If equilibration is reached before BBN, this implies an extra neutrino degree of freedom, which would be in tension with current data~\cite{Patrignani:2016xqp}. At the recombination epoch, most neutrinos have to be free streaming to agree with the temperature and polarization CMB data and this imposes additional constraints on the parameter space of the new interactions~\cite{Hannestad:2004qu, Trotta:2004ty, Bell:2005dr, Cirelli:2006kt, Friedland:2007vv, Basboll:2008fx, Cyr-Racine:2013jua, Archidiacono:2013dua, Forastieri:2015paa, Lancaster:2017ksf, Oldengott:2017fhy, Koksbang:2017rux}.

As we discussed in the previous section, the production rate of the sterile neutrinos, $\Gamma_{\nu_s}$, is equal to the product of the damping rate and the thermally averaged conversion probability. On the other hand, the Hubble expansion rate at radiation-dominated epochs, in terms of the photon temperature $T_\gamma$, is
\begin{eqnarray}
H & = & \sqrt{\frac{4 \, \pi^3 \, g_* (T_\gamma)}{45}} \frac{T_\gamma^2}{M_{pl}}  
\simeq 1.36 \times 10^{-22} \sqrt{ g_* (T_\gamma)} \, \left(\frac{T_\gamma}{{\rm MeV}}\right)^2 \, {\rm MeV} ~,
\end{eqnarray}
with the Planck mass $M_{pl} = 1.22 \times 10^{22}$~MeV. The temperature of the sterile sector (relative to the one of active neutrinos) and the number of relativistic degrees of freedom at BBN depend on the mass of the new boson $\phi$, as discussed above. Therefore, constraints from BBN data depend on $M_\phi$, too. At $T_\nu \sim 1$~MeV, for $M_\phi \geq 1$~MeV, $g_* = g_*^{\rm SM} + 2 \cdot 7/8 \cdot \xi_{\rm nr}^4 \simeq 11.06$, with $g_* ^{\rm SM} = 10.75$, and $T_s \simeq 0.649 \, T_\nu$. whereas for $M_\phi < 1$~MeV, $g_* = g_*^{\rm SM} + (2 \cdot 7/8 + 3) \cdot \xi_{\rm rel}^4 \simeq 10.97$ and $T_s \simeq 0.465 \, T_\nu$. At later times (lower temperatures), sterile neutrino recoupling would make the active and sterile neutrino temperatures equal.

\begin{figure}[t]
	\centering
	\includegraphics[width=0.9\textwidth]{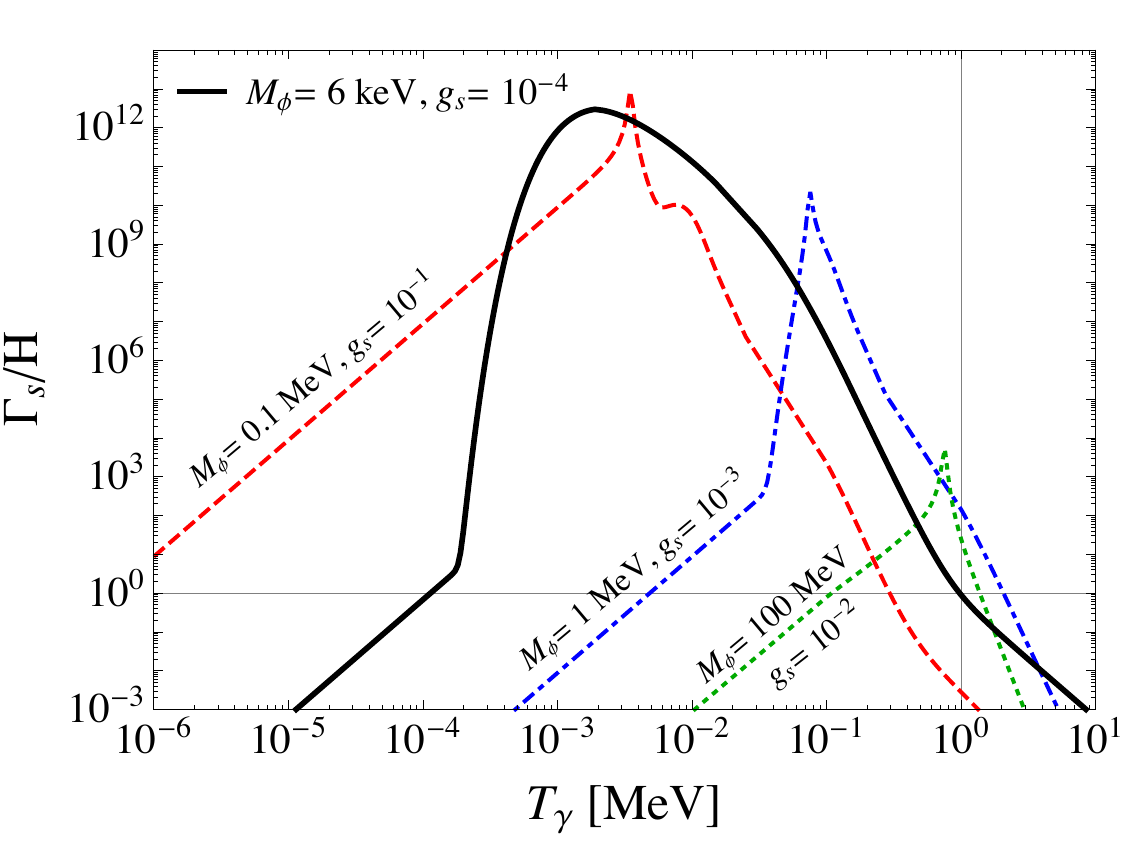}	
	\caption{The ratio of the sterile neutrino production rate, $\Gamma_{\nu_s}$, to the Hubble expansion rate, $H$, as a function of the photon temperature, $T_\gamma$, for four pairs of $(M_\phi, \, g_s)$: $(6~{\rm keV}, \, 10^{-4})$ (black solid curve), which represents our default case in Section~\ref{sec:DSNB}, $(0.1~{\rm MeV}, \, 10^{-1})$ (red dashed curve), $(1~{\rm MeV}, \, 10^{-3})$ (blue dot-dashed curve) and $(100~{\rm MeV}, \, 10^{-2})$ (green dotted curve), which are not allowed by BBN and CMB data (see text).}
	\label{fig:gsh}
\end{figure}

Before discussing in more detail the allowed regions in the parameters space of the hidden sector, $(M_\phi, \, g_s)$ (fixing $\theta_0$ and $m_s$), we illustrate in Fig. \ref{fig:gsh} how BBN and CMB cosmological constraints apply. We show the ratio of the production rate of sterile neutrinos to the Hubble expansion rate, $\Gamma_{\nu_s} /H$, as a function of the photon temperature, for several representative pairs of (excluded) values of $(M_\phi, \, g_s)$. We also show this ratio for the default values we consider in Section~\ref{sec:DSNB} (black solid curve), $M_\phi = 6$~keV and  $g_s = 10^{-4}$. In this case, $\Gamma_{\nu_s} /H < 1$ for $T_\gamma > T_{\rm BBN} = 1$~MeV and  $T_\gamma < T_{\rm CMB} \sim 1$~eV, so in these temperature regimes, sterile neutrinos are not in equilibrium with active neutrinos and both BBN and CMB constraints are satisfied. Nevertheless, for the other parameter sets presented in the figure, $\Gamma_{\nu_s} /H >1$ at either $T_{\rm BBN}$ or $T_{\rm CMB}$.  As a consequence, these parameter choices are in conflict with the constraints on the number of effective neutrino species or the condition of free-streaming at recombination (see below). In the following, we investigate the excluded/allowed regions of the $(M_\phi, \, g_s)$ parameter space in more detail.

\begin{figure}[t]
	\centering
	\includegraphics[width=\textwidth]{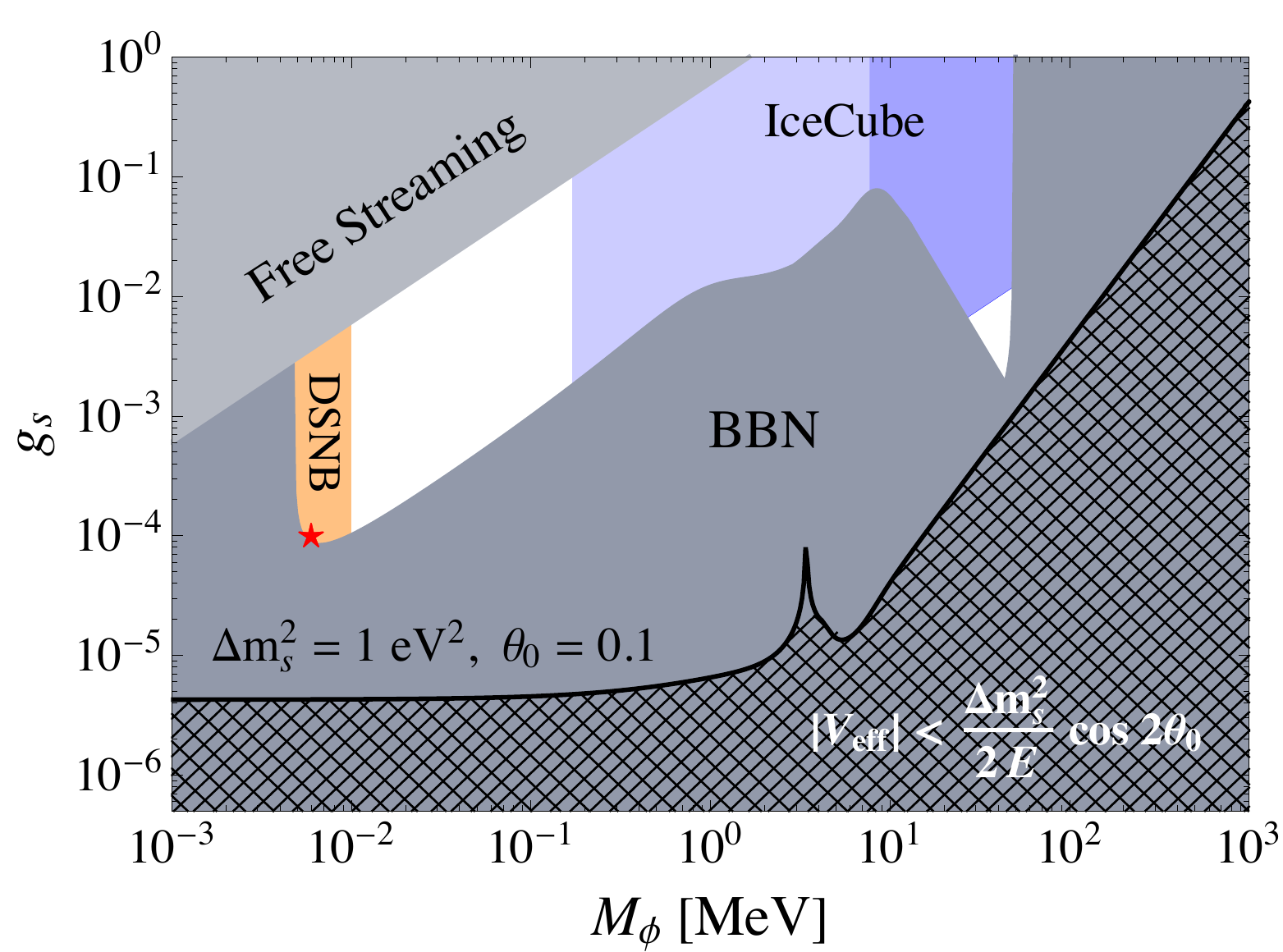}	
	\caption{Bounds on the mass of the new vector boson, $M_\phi$, and the coupling, $g_s$, of the hidden sector interactions, from Eq.~(\ref{eq:Ls}). The lower gray region shows where constraints from the effective number of neutrinos at BBN exclude ($M_\phi,g_s$).  The constraint from imposing that all three active neutrinos are free streaming at the CMB epoch are represented by the gray exclusion region in the upper left corner of the plot. The cross-hatched region corresponds to $|V_{\rm eff}| < \Delta m_s^2/(2 \, E) \cos 2 \theta_0$ at $T_\nu = 1$~MeV and using $E = 3.15 \, T_\nu$, where the standard active-sterile neutrino mixing scenario is recovered. For higher temperatures, this region moves to smaller $g_s$, c.f., Eqs.~(\ref{eq:VsOsc}) and~(\ref{eq:VsOsch}). All the results are obtained for $\Delta m_s^2 = 1$~eV$^2$ and $\theta_0 = 0.1$. We also depict the regions that can be tested by searching for potential dips in the DSNB flux in DUNE or HK (orange region, along with the default case we consider in this work, marked with a red star) and in the cosmic neutrino spectrum in IceCube (light blue region for mostly active neutrinos as targets and slightly darker blue for mostly sterile neutrinos as targets). See text for details.
	}
	\label{fig:constraints}
\end{figure}

\subsection{Big Bang Nucleosynthesis constraints}

If sterile neutrinos recouple with the SM sector before the BBN time, they would violate observational data and hence this constrains the parameter space $(M_\phi,g_s)$. The equilibration between sterile and active neutrinos would occur when the production rate of sterile neutrinos, $\Gamma_{\nu_s}(\nu_a \to \nu_s)$, exceeds the expansion rate of the Universe. The temperature when equilibration is reached is the recoupling temperature, so it is defined as $\Gamma_{\nu_s} (T_{\rm rec})  = H(T_{\rm rec})$. In order satisfy the BBN bound on $N_{\rm eff}$, we find the constrained parameter space for $(M_\phi, g_s)$ by demanding the recoupling temperature to be lower than temperatures below which BBN can be affected,
\begin{equation}
\label{eq:BBNlimit}
\Gamma_{\nu_s} (T)  < H(T) \quad {\rm for }  \quad T \geq  T _{\rm BBN} = 1~{\rm MeV} ~.
\end{equation}
At the BBN epoch, $T_\nu = T_\gamma$. Using the expressions for the interaction rates from the previous section, the resulting constraints on the coupling $g_s$, as a function of $M_\phi$, are depicted in Fig.~\ref{fig:constraints}. The excluded values of $(M_\phi, \, g_s)$ are represented by the gray region, discussed below. In the region of the parameters space where the effective potential due to sterile-sterile neutrino interactions is smaller than the vacuum term, $|V_{\rm eff}| < \Delta m_s^2/(2 \, E) \, \cos 2\theta_0$, the mixing angle is the one in vacuum. Thus, the lower the temperature, the more constraining this condition is (for high temperatures, the region moves towards large masses and couplings). For $T_\nu = 1$~MeV, this is represented by the cross-hatched area, which is excluded due to SM interactions, given the very large mixing we consider~\cite{Barbieri:1989ti, Kainulainen:1990ds, Barbieri:1990vx, Enqvist:1991qj, Shi:1993hm, Kirilova:1997sv, Abazajian:2001nj, DiBari:2001ua, Dolgov:2003sg, Melchiorri:2008gq, Hannestad:2012ky, Mirizzi:2012we, Mirizzi:2013gnd, Saviano:2013ktj, Jacques:2013xr, Hannestad:2015tea, Bridle:2016isd}.

We now turn to the BBN constraints in three limiting cases: the limit of $M_\phi \ll T_{\rm BBN}$, in the limit of $M_\phi \gg T_{\rm BBN}$, and the case in which equilibration can potentially happen before BBN.

\subsubsection {Low-mass limit ($M_\phi \ll T_{\rm BBN}$)}

When $\phi$ bosons are relativistic at the BBN epoch, i.e., $M_\phi < T_{\rm BBN}$, the temperature of the sterile sector is given by Eq.~(\ref{eq:xiA}), i.e., $T_s = 0.465 \, T_\nu$ and the effective potential (only from interactions in the sterile sector, i.e., neglecting the SM contribution) is given by, $V_{\rm eff} = g_s^2 \, T_s^2/(8 \, E)$ (see Eq.~(\ref{eq:Vefflim})). Since we approximate freeze-out as instantaneous at $T_{\rm BBN} = 1$~MeV, the neutrino and photon temperatures are approximately equal~\cite{Grohs:2017iit}. In order to obtain the constraints on the $(M_\phi, g_s)$ parameter space, it is useful to determine different regimes.  It is useful to consider the regimes such that:
\begin{eqnarray}
\label{eq:VsVSM}
V_{\rm eff, s} > |V_{\rm eff, SM}| : \hspace{5mm} g_s & > & 2.3 \times 10^{-9} \, (T_\nu/{\rm MeV})^2 ~, \\
\label{eq:VsOsc}
V_{\rm eff, s} > \Delta m_s^2 / (2 \, E) \cos 2 \theta_0 : \hspace{5mm} g_s & > & 4.3 \times 10^{-6} \, ({\rm MeV}/T_\nu)  ~, \\
\label{eq:t}
t\textrm{-channel dominance:} \hspace{5mm} g_s & > & 1.9 \, \left(M_\phi/T_\nu\right)^2 ~.
\end{eqnarray}
When the sterile neutrino cross section is $t$-channel dominated,
\begin{eqnarray}
(t\textrm{-channel}) && \nonumber \\
\label{eq:GstGSM}
\Gamma_{\rm int, s} > \Gamma_{\rm int, SM} : \hspace{5mm} g_s & > & 2.1 \times 10^{-5} \, (M_\phi/{\rm MeV})^{1/2} \, (T_\nu/{\rm MeV})^{1/2} ~, \\
\label{eq:VsGst}
V_{\rm eff, s} > \Gamma_{\rm int, s}/2  : \hspace{5mm} g_s & < & 2.4 \, \left(M_\phi/T_{\nu}\right) ~,  \\
\label{eq:GstOsc}
\Gamma_{\rm int, s}/2 > \Delta m_s^2 / (2 \, E) \cos 2 \theta_0  : \hspace{5mm} g_s & > & 3.2 \times 10^{-3} \, \left(M_\phi/{\rm MeV}\right)^{1/2} \, ({\rm MeV}/T_\nu) ~,
\end{eqnarray}
while for the $s$-channel dominated sterile neutrino cross section,
\begin{eqnarray}
(s\textrm{-channel}) && \nonumber  \\
\label{eq:GssGSM}
\Gamma_{\rm int, s} > \Gamma_{\rm int, SM} : \hspace{5mm} g_s & > & 2.3 \times 10^{-10} \, ({\rm MeV}/M_\phi) \, (T_\nu/{\rm MeV})^{3} ~, \\
\label{eq:VsGss}
V_{\rm eff, s} > \Gamma_{\rm int, s}/2  : \hspace{5mm} M_\phi & < & 1.3 \, T_{\nu} ~,  \\
\label{eq:GssOsc}
\Gamma_{\rm int, s}/2 > \Delta m_s^2 / (2 \, E) \cos 2 \theta_0  : \hspace{5mm} g_s & > & 5.5 \times 10^{-6} \, \left( {\rm MeV}/M_\phi \right) ~,
\end{eqnarray}
where we have used $\Delta m_s^2 = 1$~eV$^2$, $\theta_0 = 0.1$ and $E = 3.15 \, T_\nu$. These inequalities show that $\Gamma_{\rm int, s}$ dominates $\Gamma_{\rm int, SM}$ for most of the parameter space under discussion.

We first consider the case where the $t$-channel is the most important contribution to the sterile neutrino
interaction cross section, i.e., Eq.~(\ref{eq:t}), which roughly represents half of the low mass region ($M_\phi \lsim 1$~MeV) depicted in Fig.~\ref{fig:constraints}. Using Eq.~(\ref{eq:sig-th}), the total interaction rate in the sterile sector is given by
\begin{equation}
\label{eq:Gt}
\Gamma_{\rm int, s} \simeq \left(2 \,  \frac{g_s^4}{4 \, \pi \, M_\phi^2} \right)
\left(  \frac{3 \, \zeta(3)}{4 \, \pi^2} \, g_{\nu_s} \, T_s^3 \right) \simeq 2.9 \times 10^{-3} \, g_s^4 \, \frac{T_\nu^3}{M_\phi^2} \hspace{1cm} (t{\rm -channel}) ~,
\end{equation}
where the first parenthesis corresponds to the thermal average of the sum of cross sections for $\nu_s + \bar\nu_s \to \nu_s + \bar\nu_s$ and $\nu_s + \nu_s \to \nu_s + \nu_s$, whereas the second term is the equilibrium number density of sterile neutrinos, $n_s (T_s)$.

In the region of the parameter space where $|V_{\rm eff, s}| > \Delta m_s^2 / (2 \, E) \, \cos 2 \theta_0$ and $\Gamma_{\rm int, s}/2$ ($t$-channel), i.e.,  $4.3 \times 10^{-6} \, ({\rm MeV}/T_\nu) < g_s < 2.4 \, \left(M_\phi/T_\nu\right)$, the thermal average of the probability of active-sterile neutrino conversion can be approximated by
\begin{equation}
\langle P(\nu_a \to \nu_s) \rangle \simeq \frac{1}{2} \, \sin^2 2 \theta_m 
\simeq \frac{1}{2}
\left( \frac{\Delta m_s^2 \, \sin 2 \theta_0}{2 \, E \, V_{\rm eff, s}}\right)^2  \simeq 
8 \, \frac{\left(\Delta m_s^2 \, \sin 2 \theta_0\right)^2}{g_s^4 \, T_s^4} 
\end{equation}
and thus, from the condition in Eq.~(\ref{eq:BBNlimit}), the excluded region is given by
\begin{equation}
\label{eq:boundVt}
M_\phi < 4.7 \, {\rm keV}  \, \left( \frac{T_\nu}{\rm MeV} \right)^{-3/2} ~.
\end{equation}
Hence, in the limit in which the effective potential suppresses oscillations and the interaction rate is dominated by the $t$-channel cross section, the production rate does not depend on the coupling $g_s$, because both $V_{\rm eff}^2$ and the ($t$-channel) cross section are proportional to $g_s^{4}$. Accordingly, only constraints on $M_\phi$ can be set within this region~\cite{Cherry:2016jol}. This is represented by the vertical line in the left of Fig.~\ref{fig:constraints}, marking the boundary of the dark gray region.

In the region of the parameter space where $\Gamma_{\rm int, s}/2 > V_{\rm eff, s}$ and $\Delta m_s^2 / (2 \, E) \, \cos 2 \theta_0$, i.e., $g_s > 2.4 \, \left(M_\phi/T_\nu\right)$ and $g_s  > 3.2 \times 10^{-3} \, \left(M_\phi/{\rm MeV}\right)^{1/2} \, ({\rm MeV}/T_\nu)$, interactions interrupt oscillations; they act as a `measurement' of the neutrino state (Turing's or Zeno's paradox)~\cite{Harris:1980zi}. In this situation, the average of the conversion probability can be approximated as
\begin{equation}
\langle P(\nu_a \to \nu_s) \rangle \simeq \frac{1}{2} \, \left( \frac{\Delta m_s^2 \, \sin 2 \theta_0}{E \, \Gamma_{\rm int, s} }\right)^2 ~,
\end{equation}
which results in the excluded region
\begin{equation}
g_s \lsim 0.17 \, \left( \frac{M_\phi}{\rm MeV}\right)^{1/2} \, \left( \frac{T_\nu}{\rm MeV} \right)^{-7/4} ~.
\end{equation}
This excludes $\phi$ boson masses in the range $M_\phi < 4.7 \, {\rm keV} \, (T_\nu/{\rm MeV})^{-3/2}$ for $g_s > 2.4 \, \left(M_\phi/T_\nu\right)$, which is complementary to the exclusion region represented by Eq.~(\ref{eq:boundVt}). 

In the low-mass region where the $t$-channel dominates the total interaction rate, there is only a very small corner for which the mixing angle is approximately that of vacuum, $1.9 \, (M_\phi/T_\nu)^2 < g_s < 4.3 \times 10^{-6} \, ({\rm MeV}/T_\nu) $. In that case, $\Gamma_{\rm int, s} \gsim \Gamma_{\rm int, SM}$, and so, equilibration is even more effective than in the standard scenario of sterile neutrino production in the early Universe. Therefore, for the large mixing angle we consider, full equilibration of sterile neutrinos is achieved and BBN constraints are violated.

Next, we focus on the parameter region where the $s-$channel is the most relevant one in the interaction cross section (of the sterile sector), i.e., $g_s < 1.9 \, (M_\phi/T_\nu)^2$. It is interesting to stress, as can be seen from Fig.~\ref{fig:sigs-avg}, that as a consequence of the thermal averaging of the cross section, the $s$-channel is the most important contribution over several orders of magnitude in temperature, not just at $T_s\sim M_\phi$. Therefore, the constraints for most of the parameter space we consider are a consequence of $s$-channel interactions. In this regime, and using Eq.~(\ref{eq:sig-th}), the total interaction rate in the sterile sector is given by
\begin{equation}
\Gamma_{\rm int, s} \simeq \left(  \frac{\pi \, g_s^2 \, M_\phi^2}{18 \, T_\nu^2 \, T_s^2 \, \zeta(3)^2} \right) \, \left(  \frac{3 \, \zeta(3)}{4 \, \pi^2} \, g_{\nu_s} \, T_s^3 \right) \simeq 10^{-2} \, g_s^2 \, \frac{M_\phi^2}{T_\nu} \hspace{1cm} (s{\rm -channel}) ~,
\end{equation}

From Eqs.~(\ref{eq:GssGSM}) and~(\ref{eq:GssOsc}), when the $s$-channel is most important ($g_s < 1.9 \, (M_\phi/T_\nu)^2$) and $\phi$ bosons are relativistic at the BBN epoch, the effective potential is always larger than the damping term (in the region depicted in Fig.~\ref{fig:constraints}). Moreover, the effective potential suppresses the vacuum mixing in the range $g_s > 4.3 \times 10^{-6} \, ({\rm MeV}/T_\nu)$, Eq.~(\ref{eq:VsOsc}). Under these conditions, the excluded region is given by
\begin{equation}
\label{eq:bound-resA}
g_s < 8.8 \times 10^{-3} \, \left( \frac{M_\phi}{\rm MeV}\right) \, 
\left( \frac{T_\nu}{\rm MeV} \right)^{-7/2} ~,
\end{equation}
which approximately represents the dark gray region limited by the diagonal line in Fig.~\ref{fig:constraints}. For $M_\phi \lsim T_{\rm BBN}$, the low-mass approximation (for the interaction rate and the effective potential) is less accurate and one should use the full numerical result, which produces the shoulder at $T_\nu \sim 1$~MeV.

For $g_s < 4.3 \times 10^{-6} \, ({\rm MeV}/T_\nu)$ and $g_s < 1.9 \, (M_\phi/T_\nu)^2$, vacuum mixing is recovered. At $T_\nu = T_{\rm BBN}$ interactions in the sterile sector are more important than SM collisions between active neutrinos, Eq.~(\ref{eq:GssGSM}), in the region shown in Fig.~\ref{fig:constraints}, but the equilibration condition is not satisfied for $g_s > 2.1 \times 10^{-9} \, ({\rm MeV}/M_\phi)$ (left-bottom region in Fig.~\ref{fig:constraints}). However, in that region of the parameter space and at $T_\nu \sim $few MeV, the $t$-channel contribution is again more important than the $s$-channel one. In this case, vacuum mixing is also recovered and $\Gamma_{\rm int, s} \sim \Gamma_{\rm int, SM}$, so equilibration between the active and sterile sectors is achieved before the BBN epoch in a similar fashion as in the usually considered active-sterile neutrino mixing scenario.

\subsubsection {High-mass limit ($M_\phi \gg T_{\rm BBN}$)}

When $\phi$ bosons are non-relativistic at the BBN epoch, i.e., $M_\phi > T_{\rm BBN}$, the temperature of sterile neutrinos is given by Eq.~(\ref{eq:xiB}), i.e., $T_s = 0.649 \, T_\nu$ (and $T_\nu = T_\gamma$), and the effective potential (only from interactions in the sterile sector, i.e., neglecting the SM contribution) is $V_{\rm eff} = - \left(7 \, \pi^2 \, g_s^2/45\right) \, \left(E \, T_s^4/M_\phi^4\right)$, Eq.~(\ref{eq:Vefflim}). In this limit, both $s$- and $t$-channel contributions to the total cross section are relevant to determine constraints in different regimes. 

The $t$-channel cross section is important for large $\phi$ masses and couplings and temperatures close to $T_{\rm BBN}$, whereas the $s$-channel contribution is the dominant one to set bounds for small couplings in the entire mass interval considered ($1~{\rm MeV} < M_\phi < 1$~GeV). This can be qualitatively understood as follows. If we were to consider only interactions in the new sterile sector, due to thermal averaging, the $t$-channel would be the dominant one at $T_\nu / M_\phi < {\cal O}(100)$ (see Fig.~\ref{fig:sigs-avg}). Thus, for temperatures close to $T_{\rm BBN}$, it is more important for larger masses. Besides, for large couplings the effective potential suppresses the mixing angle in the medium, but $V_{\rm eff, s}^2 \propto g_s^4$ and $\Gamma_{\rm int, s} \propto g_s^4$, so the constraints so obtained do not depend on the coupling. For smaller couplings, mixing can be resonantly enhanced around some temperature close to BBN, so the sterile neutrino production rate would be proportional to $g_s^4$ and hence, the larger the mixing the more effective equilibration would be and the more stringent the constraints would be. This explains the upper right part in Fig.~\ref{fig:constraints}. On the other hand, the large enhancement in the interaction cross section produced by the $s$-channel contribution also results in a suppression of the mixing angle. Unlike what happens for the $t$-channel, now $\Gamma_{\rm int, s} \propto g_s^2$, so the sterile neutrino production rate would be proportional to $g_s^{-2}$ (if no SM interactions were present) and thus, the larger the coupling the smaller the impact on BBN data (equilibration is more difficult to be reached). For small $\phi$ masses, only the $s$-channel contributes to set limits on the $(M_\phi, g_s)$ parameter space and this bound smoothly connects with the low-mass case discussed above and depicted in Fig.~\ref{fig:constraints}. For $M_\phi > 100$~MeV, interactions in the new sector alone would leave an allowed region; the $t$-channel not being efficient enough and the $s$-channel suppressing too much the sterile-active mixing angle. However, in this region, SM interactions of active neutrinos take over and thermalize sterile neutrinos, excluding that part too.

Although we do not have an analytic expression for the $s$-channel contribution to the total interaction rate, it is illustrative to consider the $t$-channel dominated rate. Analogously to the low-mass limit, it is useful to determine different regimes, 
\begin{eqnarray}
\label{eq:VsVSMh}
|V_{\rm eff, s}| > |V_{\rm eff, SM}| : \hspace{5mm} g_s & > & 2.3 \times 10^{-10} \, (M_\phi/{\rm MeV})^2 ~, \\
\label{eq:VsOsch}
|V_{\rm eff, s}| > \Delta m_s^2 / (2 \, E) \cos 2 \theta_0 : \hspace{5mm} g_s & > & 4.3 \times 10^{-7} \, (M_\phi/{\rm MeV})^2 \, ({\rm MeV}/T_\nu)^3  ~, \\
\label{eq:GstGSMh}
\Gamma_{\rm int, s} > \Gamma_{\rm int, SM} : \hspace{5mm} g_s & > & 7.1 \times 10^{-6} \, (M_\phi/{\rm MeV}) ~, \\
\label{eq:VsGsth}
|V_{\rm eff, s}| > \Gamma_{\rm int, s}/2  : \hspace{5mm} g_s & < & 2.8 ~,  \\
\label{eq:GstOsch}
\Gamma_{\rm int, s}/2 > \Delta m_s^2 / (2 \, E) \cos 2 \theta_0  : \hspace{5mm} g_s & > & 1.1 \times 10^{-3} \, \left(M_\phi/{\rm MeV}\right) \, ({\rm MeV}/T_\nu)^{3/2} ~,
\end{eqnarray}
where we have also used $\Delta m_s^2 = 1$~eV$^2$, $\theta_0 = 0.1$ and $E = 3.15 \, T_\nu$.

Using Eq.~(\ref{eq:sig-th}), the total interaction rate in the sterile sector for $M_\phi \gg T_{\rm BBN}$, when it is dominated by the $t$-channel contribution, is given by
\begin{equation}
\label{eq:Gh}
\Gamma_{\rm int, s} \simeq \frac{5}{2} \, \left(  \frac{4 \, g_s^4 \, T_\nu \, T_s}{M_\phi^4 \, \zeta(3)^2} \right)
\left(  \frac{3 \, \zeta(3)}{4 \, \pi^2} \, g_{\nu_s} \, T_s^3 \right) \simeq 0.22 \, g_s^4 \, \frac{T_\nu^5}{M_\phi^4} ~.
\end{equation}

In this regime, the effective potential is always larger than the damping term\footnote{This is so even when considering the $s$-channel contribution, except for $M_\phi \sim E \simeq 3.15 \, T_\nu$, due to the behavior of the effective potential.}, Eq.~(\ref{eq:VsGsth}), and it suppresses vacuum mixing for $g_s > 4.3 \times 10^{-7} \, (M_\phi/{\rm MeV})^2 \, ({\rm MeV}/T_\nu)^3$, Eq.~(\ref{eq:VsOsch}), which results in an excluded region given by
\begin{equation}
\label{eq:boundVh}
M_\phi > 49 \, {\rm MeV} \, \left( \frac{T_\nu}{\rm MeV} \right)^{9/4} ~.
\end{equation}
For $T_\nu = T_{\rm BBN}$, this corresponds to the vertical limit of the dark gray region on the right-top part of Fig.~\ref{fig:constraints}. If we impose the equilibration condition at higher temperatures and we substitute Eq.~(\ref{eq:VsOsch}) into Eq.~(\ref{eq:boundVh}), we get
\begin{equation}
\label{eq:boundgVh}
g_s > 7.7 \times 10^{-5} \, \left( \frac{M_\phi}{\rm MeV} \right)^{2/3} ~,
\end{equation}
where we take $g_*$ to be constant within the relevant temperature range.

On the other hand, for small couplings, $g_s < 4.3 \times 10^{-7} \, (M_\phi/{\rm MeV})^2$, there is always a temperature $T_\nu > T_{\rm BBN}$, such that $|V_{\rm eff}| = \Delta m_s^2 / (2 \, E) \cos 2 \theta_0$. Consequently, mixing is not only unsuppressed, but it is resonantly enhanced before BBN, and the conversion probability is maximal (i.e., $\langle P(\nu_a \to \nu_s) \rangle =1/2$). For $7.1 \times 10^{-6} \, (M_\phi/{\rm MeV}) < g_s < 4.3 \times 10^{-7} \, (M_\phi/{\rm MeV})^2$, the excluded region is given by\footnote{Note that Eq.~(\ref{eq:GstGSMh}) implies Eq.~(\ref{eq:VsVSMh}) for $M_\phi < 1$~GeV.}
\begin{equation}
9.5 \times 10^{-6} \, \left(\frac{M_\phi}{\rm MeV}\right) \, \left(\frac{\rm MeV}{T_\nu}\right)^{3/4} < g_s <  4.3 \times 10^{-7} \, \left(\frac{M_\phi}{{\rm MeV}}\right)^2  ~,
\end{equation}
with $g_s = 4.3 \times 10^{-7} \, (M_\phi/{\rm MeV})^2 \, ({\rm MeV}/T_\nu)^3$, or equivalently,
\begin{equation}
\label{eq:boundhres}
2.7 \times 10^{-5} \, \left(\frac{M_\phi}{\rm MeV}\right)^{2/3} < g_s < 4.3 \times 10^{-7} \, \left(\frac{M_\phi}{{\rm MeV}}\right)^2 ~,
\end{equation}
which only applies\footnote{Note that the upper limit in this range of applicability for $M_\phi$ would not be present if SM interactions were not considered and Eq.~(\ref{eq:boundhres}) would apply for $M_\phi > 22$~MeV.} for $22~{\rm MeV} < M_\phi < 54$~MeV, but it is more constraining than Eq.~(\ref{eq:boundgVh}). For larger masses (and $\Gamma_{\rm int, s} > \Gamma_{\rm int, SM}$), mixing is as in vacuum and the excluded region is
\begin{equation}
\label{eq:boundh}
g_s > 2.1 \times 10^{-5} \, \left(\frac{M_\phi}{\rm MeV}\right) \, \left(\frac{\rm MeV}{T_\nu}\right)^{3/4} ~.
\end{equation}
Using also the condition of vacuum mixing, i.e., $g_s < 4.3 \times 10^{-7} \, (M_\phi/{\rm MeV})^2 \, ({\rm MeV}/T_\nu)^3$, Eq.~(\ref{eq:boundh}) reads
\begin{equation}
g_s > 7.8 \times 10^{-5} \, \left( \frac{M_\phi}{\rm MeV} \right)^{2/3}  ~,
\end{equation}
which approximately (up to a factor $(\cos 2 \theta_0)^{2/3}$) coincides with Eq.~(\ref{eq:boundgVh}).

For $2.3 \times 10^{-10} \, (M_\phi/{\rm MeV})^2 < g_s < 7.1 \times 10^{-6} \, (M_\phi/{\rm MeV})$, i.e., when $|V_{\rm eff, s}| > |V_{\rm eff, SM}|$ and $\Gamma_{\rm int, SM} > \Gamma_{\rm int, s}$, equilibration between sterile and active neutrinos is always reached before BBN, because the production rate is the product of maximal conversion probability and SM interactions. For smaller couplings, equilibration proceeds as in the usual sterile-active scenario with no extra interaction term. 

The matching from the low-mass limit constraints discussed in the previous section is explained in terms of the $s$-channel contribution from interactions in the sterile sector. For low masses, the $s$-channel contribution, for which we have no analytic expression, becomes the dominant one at temperatures around 1~MeV. For instance, for $M_\phi \simeq 10$~MeV the $s$-channel peak of the production rate occurs around $T_\nu \sim 1$~MeV when $\Gamma_{\rm int,s}/2 \sim |V_{\rm eff, s}|$. However, the suppression of the mixing angle is not enough to keep sterile neutrinos out of equilibrium down to $T_\nu = T_{\rm BBN}$ for $g_s < 7 \times 10^{-2}$. For larger masses, the resonance takes place at higher temperatures, and the production rate is larger than the expansion rate for even smaller couplings. Nevertheless, for very small coupling $g_s$, the production rate is given by SM active-active neutrino interactions and active-sterile neutrino mixing, and equilibration proceeds as in the usual sterile neutrino scenario with no extra interactions.

\subsection{Free streaming in the Cosmic Microwave Background epoch}

Within the standard scenario, active neutrinos start free streaming well before the CMB epoch, generating anisotropic stress, which results in baryon acoustic peaks to be suppressed. However, if neutrinos were interacting instead, the amplitude of CMB fluctuations on all sub-horizon scales at the decoupling time would be enhanced~\cite{Hannestad:2004qu, Trotta:2004ty, Bell:2005dr, Cirelli:2006kt, Friedland:2007vv, Basboll:2008fx, Cyr-Racine:2013jua, Archidiacono:2013dua, Forastieri:2015paa, Lancaster:2017ksf, Oldengott:2017fhy, Koksbang:2017rux}. In order to satisfy the free-streaming condition, we conservatively impose the sterile neutrino production rate not to be greater than the expansion rate at the CMB time ($T _\gamma \sim 1$~eV). At the epoch of recombination (even earlier, in general), the effective potential and the damping term can be neglected, and hence active-sterile neutrino mixing occurs as in vacuum. As the sterile neutrino production rate (via new interactions) scales with $T_\nu^5$ and the Universe expansion rate with $T_\gamma^2$, sterile neutrinos recouple with active ones before the CMB epoch (but they must do it after BBN). Moreover, to satisfy the free-streaming condition, they must also decouple before the CMB epoch. 

After recoupling takes place, the vector boson $\phi$ and the sterile and active neutrinos acquire a common temperature. Assuming all three active neutrino species recouple with the sterile sector, equilibrium is dictated by detailed balance and the common temperature $T_{\rm rec}$ after this occurs is
\begin{equation}
T_{\rm rec} =  \left(\frac{\left(3 + 2 \cdot 7/8\right) \, \xi_{\rm rel}^4 + 3 \cdot 2 \cdot 7/8}{3 + 4 \cdot 2 \cdot 7/8}\right)^{1/4} \, T_\nu \simeq 0.860 \, T_\nu ~, 
\end{equation}
where $T_\nu$ is the active neutrino temperature. However, before decoupling takes place, the $\phi$ bosons would decay away releasing entropy to the system and thus, the final temperature of sterile neutrinos (common to active neutrinos) when decoupled would be
\begin{equation}
T_{as} =  \left(\frac{3 + 4 \cdot 2 \cdot 7/8}{4 \cdot 2 \cdot 7/8}\right)^{1/3} T_{\rm rec} \simeq 0.969 \, T_\nu \equiv \xi_{\rm eV} \, T_\nu ~,
\end{equation}

Therefore, the region of the parameter space where sterile neutrinos have already decoupled (after recoupling) at $T_\gamma \sim 1$~eV is obtained using Eq.~(\ref{eq:BBNlimit}) at that temperature. Thus, using Eq.~(\ref{eq:Gh}) (with the substitution $T_{s} \to T_{as}$), the (complementary) excluded region is given by
\begin{equation}
\label{eq:boundfs}
g_s \gsim  0.6 \, \left( \frac{M_\phi}{\rm MeV} \right) \, \left(\frac{T_\gamma}{{\rm eV}}\right)^{-3/4} ~, 
\end{equation}
where we have used $T_\nu = (4/11)^{1/3} \, T_\gamma$ and $g_*(T_\gamma \simeq 1~{\rm eV}) = g_*^{\rm SM} ({\rm eV}) + 2 \cdot 7/8 \cdot \xi_{\rm eV}^4 \, (T_\nu/T_\gamma)^4 \simeq 3.76$, with $g_*^{\rm SM} ({\rm eV})\simeq 3.36$. This result constrains the upper-left corner (shaded with gray) in Fig.~\ref{fig:constraints}. It is interesting to note that this rough estimate is in good agreement with the bounds obtained from current CMB data\footnote{Using $G_{\rm eff}$ (or $G_\nu$) in Refs.~\cite{Lancaster:2017ksf, Oldengott:2017fhy} to be $G_{\rm eff} \equiv (\sin^2 2\theta_0/2)^{1/2} \, g_s^2/M_\phi^2$, in our notation.}, which allow a region around $g_s \sim 0.5 \, \left(M_\phi/{\rm MeV}\right)$~\cite{Lancaster:2017ksf, Oldengott:2017fhy}. 

One could wonder whether the presence of four neutrinos with similar temperatures at the epoch of recombination would be in tension with the value of $N_{\rm eff}$ obtained from Planck data~\cite{Ade:2015xua}. However, given that the (mostly) sterile neutrino becomes semi-relativistic at that time, $N_{\rm eff} < 3$ in this scenario~\cite{Mirizzi:2014ama, Cherry:2014xra, Chu:2015ipa}. This means the $N_{\rm eff}$ is in agreement with the Planck limits. Note that to properly test this region a full analysis of CMB data, including sterile neutrinos with secret interactions~\cite{Forastieri:2017oma} and without assuming contact interactions, would be required. Nevertheless, this is beyond the scope of this work.

\section{Diffuse Supernova Neutrino Background}
\label{sec:DSNB}

\subsection{Spectrum of the DSNB}

To evaluate the impact of relic sterile neutrinos on the propagation of supernova neutrinos, an estimate of the spectrum of the DSNB flux is required. Neutrinos with energies in the range of tens of MeV are copiously produced after the explosions of core-collapse supernovae (SN) of type II, Ib or Ic. Whereas a SN explosion within our galaxy would give rise to thousands (or even millions) of neutrino-induced events in current or future detectors~\cite{Ikeda:2007sa, Wurm:2011zn, Machado:2012ee, Djurcic:2015vqa, Acciarri:2016crz, HK}, this might not happen in the next decades. Nevertheless, neutrinos from all SN throughout the history of the Universe are a guaranteed flux, which is known as the DSNB or supernova relic neutrinos (see, e.g., Refs.~\cite{Beacom:2010kk, Lunardini:2010ab, Mirizzi:2015eza} for reviews). The spectrum of the DSNB flux that can be observed at Earth depends on the SN formation rate, $R_{\rm SN} (z)$, and the neutrino energy spectrum from a generic SN explosion, $dN/dE_\nu$. In the absence of neutrino absorption and without taking into account oscillations, the differential DSNB flux of flavor $a$ is formulated as 
\begin{equation}
\label{eq:flux-dsnb}
F_a(E_\nu) = \int_{0}^{z_{\rm max}} dz \, R_{\rm SN} (z) \, \frac{dN_a (E'_\nu)}{dE'_\nu} \, 
(1+z) \, \left|\frac{dt}{dz} \right|  ~,
\end{equation}
where $E'_\nu = E_\nu \, (1 + z)$ is the energy of the emitted neutrinos at redshift $z$.  We take $z_{\rm max} = 6$ (although it is not very sensitive to the exact value of $z_{\rm max}$, given that the largest contributions come from $z \lesssim 2$). The factor $dz/dt$ is given by 
\begin{equation}
\label{eq:dzdt}
\frac{dt}{dz} = - \left( H_0 (1+z) \sqrt{\Omega_m (1+z)^3 + \Omega_\Lambda}\right)^{-1} ~,
\end{equation}
with the matter density $\Omega_m = 0.308 \pm 0.012$, the dark energy density $\Omega_\Lambda = 0.692 \pm 0.012$ and the Hubble parameter $H_0 = (67.8 \pm 0.9) \, {\rm km \ s^{-1}  Mpc^{-1}}$~\cite{Ade:2015xua}.

\begin{figure} [t]
	\centering
	\includegraphics[width=0.9\textwidth]{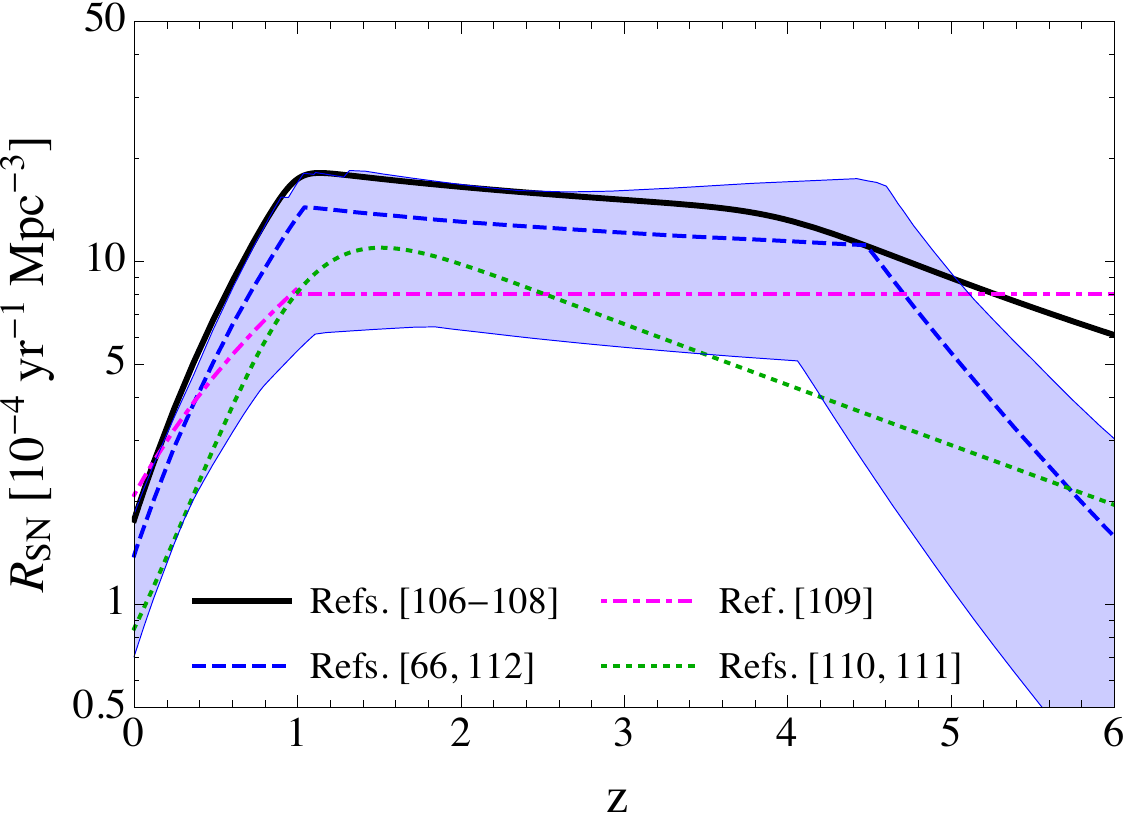}
	\caption{The cosmic SNR, using the parametrization of Ref.~\cite{Yuksel:2008cu} with the updated parameters of Ref.~\cite{Kistler:2013jza} and the conversion between SFR and SNR from Ref.~\cite{Horiuchi:2011zz} (black solid curve), which we use as our default SNR in this work. We also show other SNR parameterization: piecewise parameterization with a modified Salpeter initial mass function from Ref.~\cite{Hopkins:2006bw} (magenta dot-dashed curve), and the parameterizations used in Refs.~\cite{Porciani:2000ag, Ando:2004hc} (green dotted curve) and~\cite{Strigari:2003ig, Goldberg:2005yw} (blue dashed curve). The uncertainty band corresponds to the $3\sigma$ CL regions for the piecewise parameterization of the SFR in  Ref.~\cite{Hopkins:2006bw}, for $T_{\bar{\nu_e}} = (4-8)$~MeV, using the appropriate scaling factors.}
	\label{fig:SNR}
\end{figure}

The cosmic SN rate (SNR) is expected to be proportional to the star formation rate (SFR) and the relative normalization provides information on the frequency of optically dim core-collapsed SN~\cite{Horiuchi:2011zz}. In this work, we consider the parameterization of the SFR proposed in Ref.~\cite{Yuksel:2008cu},
\begin{equation}
\label{eq:sfr}
\dot\rho_* (z) = \dot{\rho_0} \, \left[ (1+z)^{a \, \zeta} + \left( \frac{1+z}{B} \right)^{b \, \zeta} + \left( \frac{1+z}{C}\right)^{c \, \zeta} \right]^{1/\zeta} ~,
\end{equation}
where $\dot{\rho_0} = 0.02 \, M_{\odot} {\rm yr^{-1} Mpc^{-3}}$, $a = 3.4$, $b = - 0.3$, $c = - 2.5$, $\zeta = - 10$, $B = (1 + z_1)^{1 - a/b}$ and $C = (1 + z_1)^{(b - a)/c} \, (1 + z_2)^{1 - b/c}$ for $z_1 = 1$ and $z_2 = 4$~\cite{Kistler:2013jza}.

The conversion from SFR to SNR is obtained by assuming canonical parameters for optically luminous core-collapsed SN ($M_{\rm min} = 8~M_\odot$ and $M_{\rm max} = 40 \, M_\odot$) and using a Salpeter initial mass function, defined in the range $0.1-100~M_\odot$, so that SNR per comoving volume at redshift $z$ can be written as~\cite{Horiuchi:2011zz}
\begin{equation}
\label{eq:snr}
R_{\rm SN} (z) = \frac{0.0088}{M_{\odot}} \, \dot\rho_*(z) ~.
\end{equation}
This is the default parameterization we use for the SNR throughout this work.\footnote{Although a discrepancy of approximately a factor of two between the SNR estimated from the SFR and the observed SNR was noticed~\cite{Mannucci:2007ex, Botticella:2007er, Horiuchi:2011zz}, this problem seems to have been alleviated in the last few years with further observations~\cite{Li:2010kd, Graur:2011cv, Botticella:2011nd, Mattila:2012zr, Melinder:2012nv, Dahlen:2012cm, Taylor:2014rlo, Graur:2014bua, Cappellaro:2015qia, Strolger:2015kra, Petrushevska:2016kie}.}  We show this parameterization in Fig.~\ref{fig:SNR}, along with others: the piecewise parameterization given in Ref.~\cite{Hopkins:2006bw} using a modified Salpeter initial mass function and the ones used in Refs.~\cite{Strigari:2003ig, Goldberg:2005yw} and~\cite{Porciani:2000ag, Ando:2004hc}. We also show the $3\sigma$ confidence level (CL) uncertainty band on the SFR obtained in Ref.~\cite{Hopkins:2006bw}.

For the energy spectrum of neutrinos of flavor $a$ emitted from a typical SN we consider the parametrization proposed in Ref.~\cite{Keil:2002in},
\begin{equation}
\label{eq:dnde}
\frac{dN_a}{dE_\nu} (E_\nu) \equiv F_{a}^0 (E_\nu) = \frac{L_a}{\overline{E}_a} \, \frac{(1 + \alpha_a)^{1 + \alpha_a}}{\Gamma(1 + \alpha_a) \, \overline{E}_a} 
\left( \frac{E_\nu}{\overline{E}_a} \right)^{\alpha_a} e^{- (1 + \alpha_a) E_\nu/\overline{E}_a} ,
\end{equation}
where $\overline{E}_a$ is the average neutrino energy, $L_a$ is the total energy released in neutrinos of  flavor $a$ and $\beta_a$ controls the shape of the spectrum. The values of these parameters have been studied and updated by the several groups~\cite{Totani:1997vj, Keil:2002in, Thompson:2002mw, Sumiyoshi:2006id, Marek:2007gr, Ott:2008jb, Fischer:2009af,  Huedepohl:2009wh, OConnor:2012bsj, Nakazato:2012qf, Tamborra:2012ac, Nakazato:2013maa, Summa:2015nyk, Horiuchi:2017qja}. In Ref.~\cite{Lunardini:2010ab} three parameter sets, which loosely cover the results from simulations, are considered. We take two of these representative sets, for the high (hot) and low (cold) energy cases, which embed the range of the expected fluxes. The values of these three sets are indicated in Tab.~\ref{table:dndeparam}, where $\nu_x$ denotes all the non-electron neutrinos and antineutrinos, i.e., $\nu_\mu \, (\bar{\nu}_\mu)$ and $\nu_\tau \, (\bar{\nu}_\tau)$. We assume energy is equipartitioned among the three neutrino flavors, i.e., $L_a = 5 \times 10^{52}$~ergs.

\begin{table}[t]
	\vskip 0.35in
	\begin{center}
		\begin{tabular}{|c|c|c|c|c|c|c|c|}
			\hline
			Model & $\overline{E}_{\nu_e}$ [MeV] & $\overline{E}_{\bar{\nu}_e}$  [MeV]  &  $\overline{E}_{\nu_x}$  [MeV] & $\alpha_{\nu_e}$ & $\alpha_{\bar{\nu}_e}$& $\alpha_{\nu_x}$\\
			\hline
			\hline
			HE & 12 & 15 &  18 & 3 & 3 & 2 \\
			\hline
			LE & 9 & 11 &  13 & 3 & 3  & 2 \\
			\hline
		\end{tabular}
	\end{center}
	\caption{Parameters for the neutrino spectra from core-collapsed SN, using Eq.~(\ref{eq:dnde}). The parameter sets labeled as HE (high energy) and LE (low energy) are taken from Ref.~\cite{Lunardini:2010ab}, which correspond to their hot and cold scenarios, respectively. The luminosities for the different flavors for both models are taken to be $L_{\nu_e} = L_{\bar\nu_e} = L_{\nu_x} = 5 \times 10^{52}$~ergs.
	}
	\label{table:dndeparam}
\end{table}

\begin{table}[t]
	\vskip 0.35in
	\begin{center}
		\begin{tabular}{|c|c|c|c|c|c|}
			\hline
			\multirow{2}{*}{NH} & $\nu$ & $F_{\nu_1}^0 = F_{\nu_x}^0$ & $F_{\nu_2}^0 = F_{\nu_x}^0$  
			& $F_{\nu_3}^0 = F_{\nu_s}^0$ & $F_{\nu_4}^0 = F_{\nu_e}^0$ \\ 
			\cline{2-6}
			& $\bar\nu$ & $F_{\bar{\nu}_1}^0 = F_{\bar{\nu}_e}^0$ & $F_{\bar{\nu}_2}^0 = F_{\nu_s}^0$  & $F_{\bar{\nu}_3}^0 = F_{\nu_x}^0$  & $F_{\bar{\nu}_4}^0 = F_{\nu_x}^0$\\
			\hline
			\hline
			\multirow{2}{*}{IH} & $\nu$ & $F_{\nu_1}^0 = F_{\nu_x}^0$ & $F_{\nu_2}^0 = F_{\nu_s}^0$ & $F_{\nu_3}^0 = F_{\nu_x}^0$  & $F_{\nu_4}^0 = F_{\nu_e}^0$ \\ 
			\cline{2-6}
			& $\bar\nu$ & $F_{\bar{\nu}_1}^0 = F_{\nu_s}^0$ & $F_{\bar{\nu}_2}^0 = F_{\nu_x}^0$  
			& $F_{\bar{\nu}_3}^0 = F_{\bar{\nu}_e}^0$  & $F_{\bar{\nu}_4}^0 = F_{\nu_x}^0$ \\
			\hline
		\end{tabular}
	\end{center}
	\caption{The SN neutrino flux of mass eigenstates in terms of the flux in terms of flavor eigenstates at production for the normal hierarchy (NH) and inverse hierarchy (IH) of SM neutrino masses~\cite{Esmaili:2014gya}. We assume all resonances are adiabatic and all mixing angles are different from zero. We have set the initial flux of sterile neutrinos to zero, $F_{\nu_s}^0 = 0$.} 
	\label{table:dnde-mf}
\end{table}

For the measured neutrino mixing parameters, the propagation of neutrinos produced in the interior of the core-collapsed SN proceeds in an adiabatic way~\cite{Dighe:1999bi}, i.e., initially produced neutrino mass eigenstates (due to the high densities in the center of the star) remain in such a state while the flavor composition changes along the trajectory due to the varying density.\footnote{Except from small corrections on the DSNB due to collective effects that are partly washed out due to smearing over time and over the SN population~\cite{Lunardini:2012ne}, at the surface of the star, neutrino fluxes in terms of flavor eigenstates can be expressed as a linear combination of the fluxes at production.} Even in the case of a scenario with an extra sterile neutrino, for the parameters considered in this work, adiabatic propagation also takes place (see Ref.~\cite{Esmaili:2014gya}). However, in such a case the relations among the flavor fluxes at production and the fluxes in the mass basis when exiting the star are not the same as in the standard scenario with three active neutrinos. They are indicated in Tab.~\ref{table:dnde-mf} for both SM mass hierarchies, the normal hierarchy (NH) and inverse hierarchy (IH). Finally, the redshift-integrated spectrum of neutrinos of flavor $a$ that arrive at Earth from core-collapsed SN is given by
\begin{equation}
\label{eq:nufluxEarth}
F_a (E_\nu) = \sum_{i=1}^4 |U_{ai}|^2 \, F_i(E_\nu) = \sum_{i = 1}^4 |U_{ai}|^2 \, \int_{0}^{z_{\rm max}} dz \, R_{\rm SN} (z) \, F_i^0(E') \, (1+z) \, \left|\frac{dt}{dz} \right|  ~.
\end{equation}
For the mixing parameters of active neutrinos, we use $\sin^2\theta_{12} = 0.307$ ($0.307$) and $\sin^2\theta_{23} = 0.538$ ($0.554$), $\sin^2\theta_{13} = 0.02206$ ($0.02227$) for NH (IH)~\cite{Esteban:2016qun, nufit32} (see also Refs.~\cite{Capozzi:2017ipn, deSalas:2017kay}). For the mixings of the sterile sector with active neutrinos, we use $\theta_{14} = \theta_{24} = \theta_{34} \equiv \theta_0 = 0.1$.  We set all CP violating phases to zero.

\begin{figure} [t]
	\centering
	\includegraphics[width=0.49\textwidth]{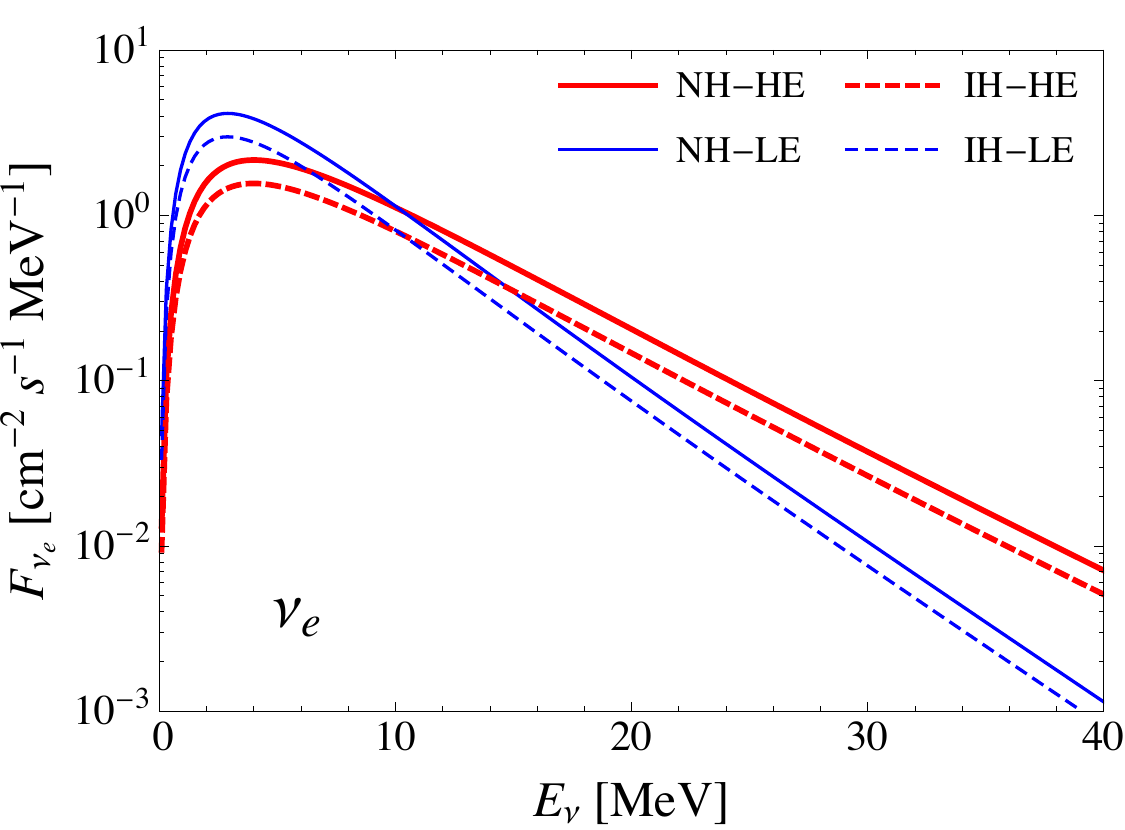}
	\includegraphics[width=0.49\textwidth]{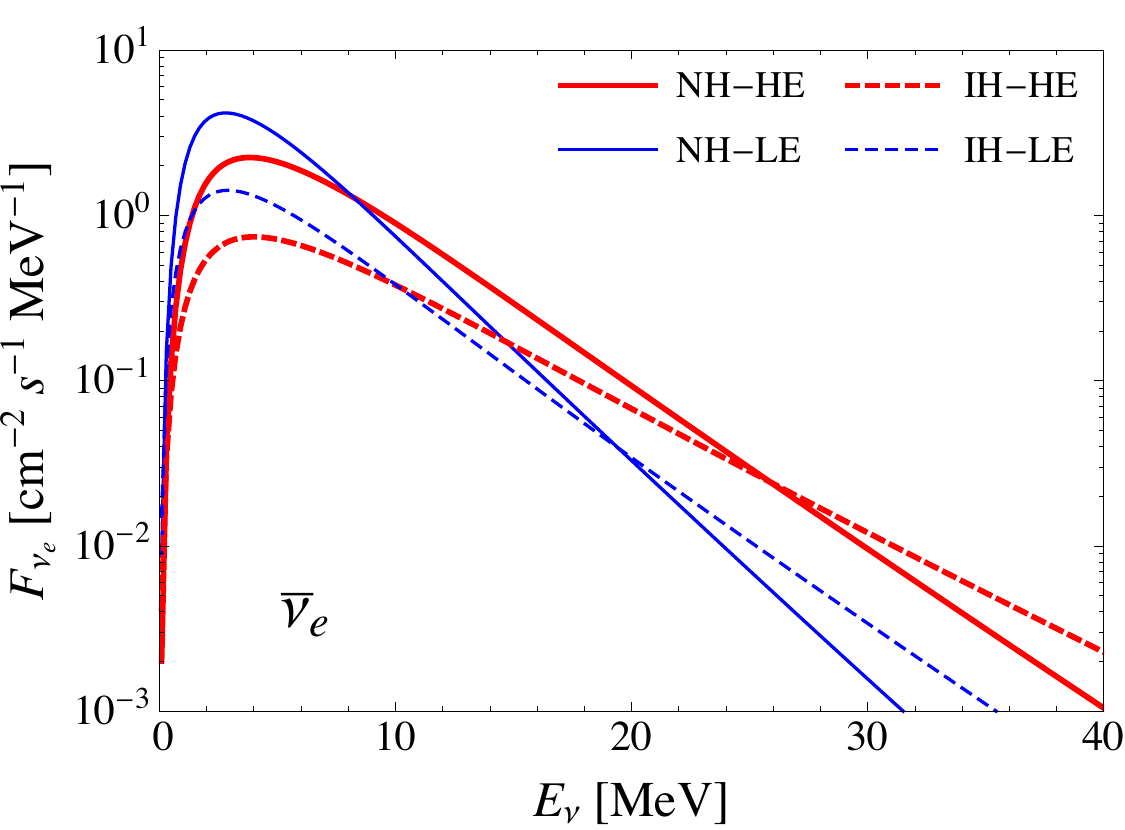}
	\caption{The $\nu_e$ (left panel) and $\bar{\nu}_e$ (right panel) DSNB fluxes for NH (solid curves) and IH (dashed curves), with the SNR  from Refs.~\cite{Yuksel:2008cu, Horiuchi:2011zz, Kistler:2013jza}, for the HE (red curves) and LE (blue curves) initial SN spectra~\cite{Lunardini:2010ab}. See text for details.}
	\label{fig:fnue0res-morder}
\end{figure}

In Fig.~\ref{fig:fnue0res-morder}, we present the $\nu_e$ and $\bar\nu_e$ DSNB spectra for NH and IH, for two of the parameter sets of the SN energy spectra (HE and LE)~\cite{Lunardini:2010ab}, as indicated in Tab.~\ref{table:dndeparam} and for the SNR from Refs.~\cite{Yuksel:2008cu, Horiuchi:2011zz, Kistler:2013jza}. Unlike what happens for the standard three-neutrino scenario, where the flux is higher for neutrinos for the IH and for antineutrinos for the NH, in the case of one extra sterile neutrino, the NH is always the case for which a larger flux is expected. Indeed, except for the $\nu_e$ flux and the NH, there is a suppression of the expected flux with respect to the three-neutrino scenario, which can be as large as a factor of about three for $\bar\nu_e$ and the IH.

\subsection{Effects of secret interactions on the DSNB spectrum}

In this section, we evaluate the DSNB absorption features due to the relic sterile neutrino background. Following the discussion from the previous section where we described the bounds from BBN and CMB observables, for the allowed region in the $(M_\phi, g_s)$ plane, sterile neutrinos would recouple with active neutrinos, then decouple, before the CMB epoch, so that most active neutrinos are free streaming by then. Assuming that after recoupling all three active neutrinos equilibrate with the sterile species, the common temperature of the system after $\phi$ bosons have decayed would be $T_{as} = 0.969 \, T_\nu$, with $T_\nu = (4/11)^{1/3} \, T_\gamma$.

Given that decoupling would have occurred when sterile neutrinos were still relativistic, there would be a relic density of (non-relativistic) sterile neutrinos at present times equal to that of a relativistic particle with temperature $T_{as}$. In this case, neutrinos from the DSNB flux could interact with the non-relativistic (mostly sterile) mass eigenstate with mass $m \simeq 1$~eV on their way to the Earth. The $s$-channel component of the $\nu_a$ and $\bar\nu_a$ interaction on the background $\nu_s$ and $\bar\nu_s$, given its resonance behavior, could have a very strong impact on the DSNB spectrum, giving rise to the absorption of part of it. On the other hand, the effect of the $t$-channel (including the elastic channel) is expected to be much smaller, so we neglect it in what follows. Indeed, for $g_s \lsim 10^{-3}$ and $M_\phi \sim 6$~keV, the optical depth from the non-resonant part of the interaction is much smaller than unity. We note that the resonant interaction would produce an on-shell $\phi$ boson, which would later decay into two sterile neutrinos. The repopulation of the active neutrino flux at lower energies would be suppressed by a factor $\sin^2 2\theta_0$. Thus, for the purpose of this paper, we can safely neglect this effect as well, the corrections being at the percent level.

Therefore, the fraction of the DSNB flux which is not absorbed due to the resonance interaction can be obtained by including the probability for a neutrino of mass eigenstate $i$ not to interact at redshift $z$ in the redshift-integrated differential DSNB flux, Eq.~(\ref{eq:nufluxEarth}), i.e.,
\begin{equation}
\label{eq:nufluxEarth-dsnb-res}
F_a (E_\nu) =  \sum_{i = 1}^4 \left| U_{ai}\right|^2 \, \int_{0}^{z_{\rm max}} dz \, P_i(E_\nu , z) \, R_{\rm SN} (z) \, F_i^0(E') \, (1+z) \, \left|\frac{dt}{dz} \right|  ~,
\end{equation}
where the probability $P_i(E_\nu , z) = e^{- \tau_i (E_\nu, z)}$ is defined in terms of the optical depth,
\begin{equation}
\label{eq:tau}
\tau_i (E_\nu, z) \simeq \int_0^z \frac{dz'}{H(z') (1+z')} \, n_s^0 \, (1 + z')^3 \, |U_{si}|^2 \, \sigma_s (z',E_\nu) \  .
\end{equation}
The factor $|U_{si}|^2$ selects the sterile component of the mass eigenstate $i$ and, as done throughout the paper, we set $\theta_0 = 0.1$. Here, $n^0_s$ denotes the present number density of sterile neutrinos (or antineutrinos), which is given by
\begin{equation}
n^0_s \simeq \frac{1}{2} \, n^0_{\nu + \bar{\nu}} \, \left(\frac{T_{as}}{T_\nu}\right)^3 \simeq 51~{\rm cm^{-3}} ~,
\end{equation}
where $n^0_{\nu + \bar{\nu}} \simeq 112 \, {\rm cm^{-3}}$ is the present active neutrino (plus antineutrino) number density per flavor. 

In our evaluation of the optical depth, we only include resonant interactions of mostly active with mostly sterile neutrinos with definite masses. Note that if the target states are the mostly active ones, i.e., $m \lesssim 0.1$~eV, in order to produce observable dips in the DSNB flux, the mass of the $\phi$ boson would have to lie in a range excluded either by BBN or by the free-streaming condition. For the parameters we consider, the resonant energies for light mass eigenstates as targets would be larger than $100$~MeV, as discussed below.

The cross section for the resonance interaction and the $\phi$ decay width appear in Eq.~(\ref{eq:sigs}). The resonant energy is given by
\begin{equation}
\label{eq:resE}
E_{\rm res} = \frac{M_\phi^2}{2 \, m_s} = 18~{\rm MeV} \, \left(\frac{M_\phi}{6~{\rm keV}}\right)^2 \, \left(\frac{1~{\rm eV}}{m_s}\right) ~,
\end{equation}
with $m_s$ the mass of the (mostly) sterile neutrino. For small couplings, the $\phi$ decay width $\Gamma_\phi \ll M_\phi$, so the cross section can be rewritten using the narrow width approximation (NWA) with $s = 2 \, m_s \, E_\nu \, (1 + z')$,
\begin{equation}
\sigma_s (z',E_\nu) \simeq 2 \, \pi \, g_s^2 \, \frac{s}{M_\phi^2} \, \delta (s - M_\phi^2) = 2 \, \pi \, g_s^2 \, \frac{(1 + z')}{M_\phi^2} \, \delta \left( (1 + z') - \frac{M_\phi^2} {2 \, m_s \, E_\nu} \right) ~.
\end{equation}
With this approximation, the integral over redshift in the optical depth, Eq.~(\ref{eq:tau}), can be analytically  performed
\begin{eqnarray}
\label{eq:tauNWA}
\tau_i (E_\nu,z) & \simeq & \frac{1}{H \left((E_{\rm res}/E_\nu) - 1\right)} \,
n_s^0 \, \left(\frac{E_{\rm res}}{E_\nu}\right)^3 \, |U_{si}|^2 \, \frac{2 \, \pi \, g_s^2}{M_\phi^2} \nonumber \\
& \simeq &  4.7 \times 10^{11} \, g_s^2 \, \left(\frac{6~{\rm keV}}{M_\phi}\right)^2 \, \left(\frac{E_{\rm res}}{E_\nu}\right)^3 \, \frac{1}{\left(\Omega_\Lambda + \Omega_m \, (E_{\rm res}/E_\nu)^3\right)} \, \left(\frac{|U_{si}|^2}{10^{-2}}\right) ~,
\end{eqnarray}
for $E_{\rm res}/(1 + z) < E_\nu < E_{\rm res}$. 

If we define the absorption factor as $f_{\rm abs} \equiv 1 - P_i(E_\nu, z)$ and take\footnote{Note that for interactions of mostly-sterile massive neutrinos ($\nu_4$), the optical depth is only suppressed by $|U_{s4}|^2 \simeq 1$. However, their contribution to the $\nu_e$ or $\bar\nu_e$ DSNB fluxes is suppressed by $|U_{e4}|^2 \simeq 0.01$ and hence, we can neglect it.} $|U_{si}|^2 = 10^{-2}$, an absorption of 10\% is reached at $E_\nu = E_{\rm res}$ for a coupling $\sim 5 \times 10^{-7}$, for $M_\phi = 6$~keV. The absorption reaches $\sim$ 99\% at $g_s \sim 3 \times 10^{-6}$, which is also excluded, as shown in Fig.~\ref{fig:constraints}.

For the range of allowed couplings presented in Fig.~\ref{fig:constraints} and $5~{\rm keV} \lesssim M_\phi \lesssim 10$~keV (such that $10~{\rm MeV} \lsim E_{\rm res} \lsim 50~{\rm MeV}$), the resulting spectra is maximally attenuated by resonant interactions, so its shape does not actually depend on the particular value of the coupling. Thus, for definiteness we take $g_s = 10^{-4}$. In Fig.~\ref{fig:constraints}, we highlight in orange the region that could be tested by measuring the DSNB with future neutrino detectors.

Our focus here is on $M_\phi$ in the keV-mass scale. In Fig.~\ref{fig:constraints} we also show in blue a different region of the parameter space that could be tested with the astrophysical neutrino flux in the TeV-PeV range. Within this type of scenario, it has been noted~\cite{Cherry:2014xra, Cherry:2016jol} that dips, similar to the ones here discussed but at higher energies, would be produced for $0.1~{\rm MeV} \lesssim M_\phi \lesssim 100$~MeV (see also Refs.~\cite{PalomaresWeiler, Weilertalk, PalomaresRuiztalk07, PalomaresRuiztalk11, Hooper:2007jr, Ioka:2014kca, Ng:2014pca} for earlier very related studies). We obtain that region by imposing $\tau (z = 0) > 1$, such that $30~{\rm TeV} < E_{\rm res} < 2$~PeV, for $|U_{s i}|^2 = 0.01$. If the target particles are (mostly) sterile neutrinos, from Eq.~(\ref{eq:tauNWA}), we get the darker blue region, whereas the lighter blue region results from considering as targets (mostly) active neutrinos with mass in the range\footnote{The lower limit, $3.15 \, T_\nu^0$ with $T_\nu^0$ the present active relic neutrino temperature, guarantees that the targets are approximately at rest in present times so that the absorption feature would not be too broad.} $3.15 \, T_\nu^0 < m_a < 0.1$~eV, requiring an extra $|U_{si}|^2$ factor for the interactions of mostly-active massive neutrinos.

\begin{figure}[t]
	\centering
	\includegraphics[width=0.49\textwidth]{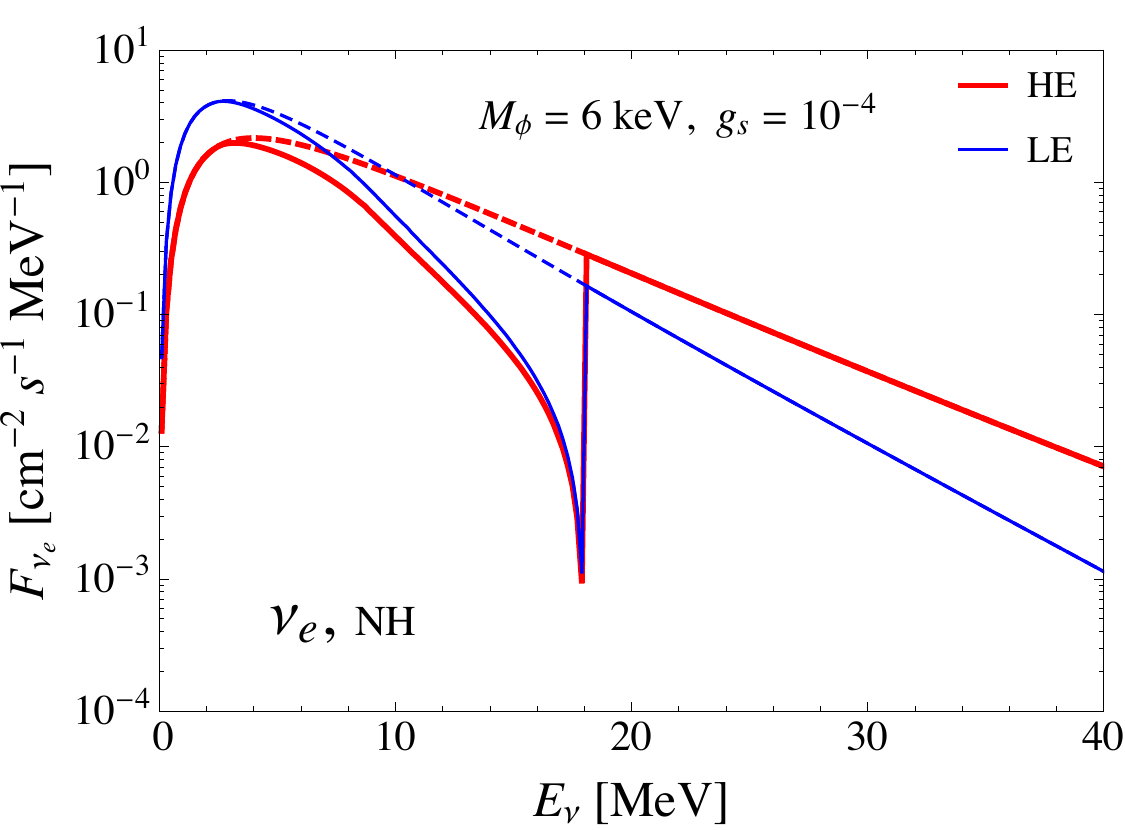}
	\includegraphics[width=0.49\textwidth]{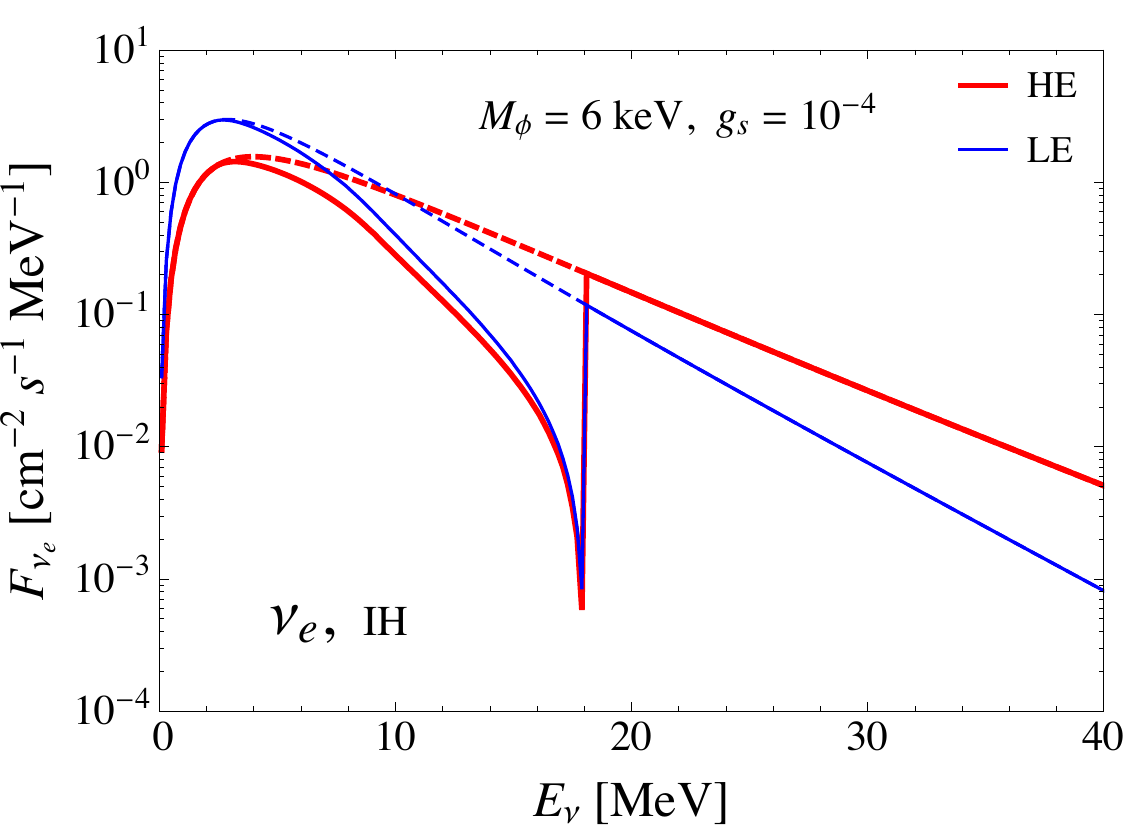}
	\includegraphics[width=0.49\textwidth]{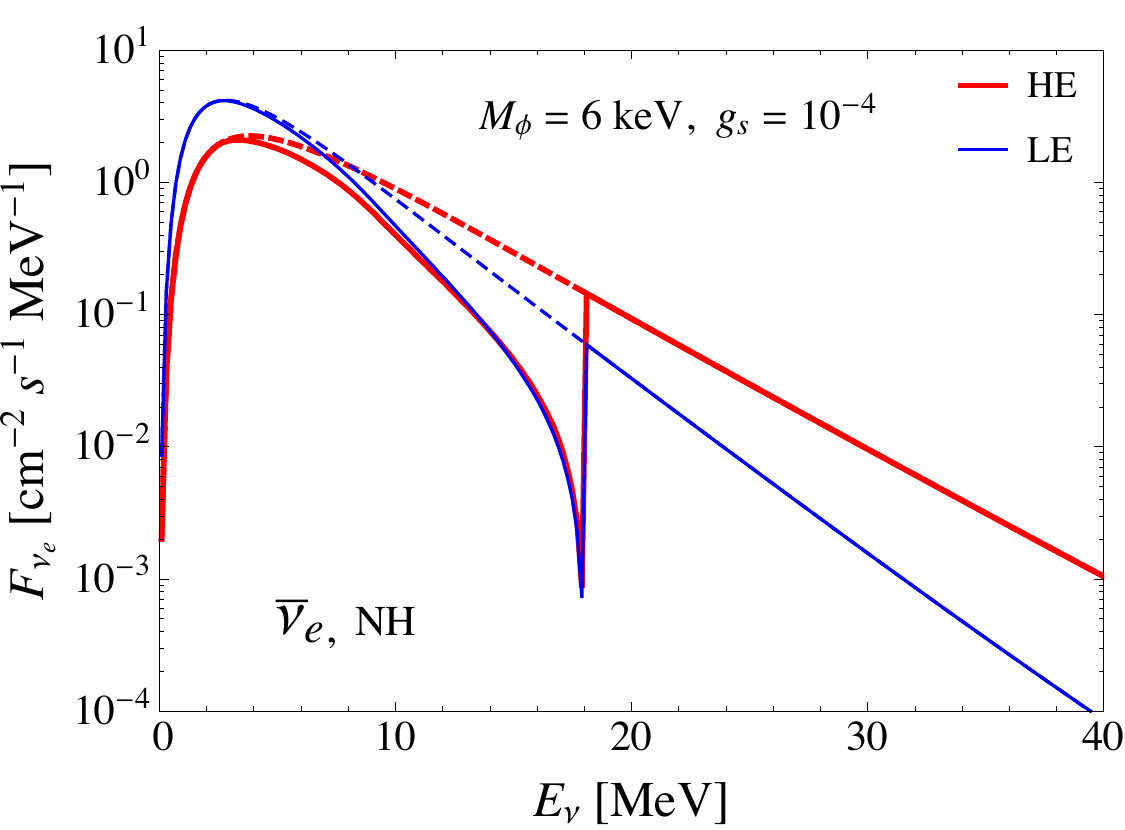}
	\includegraphics[width=0.49\textwidth]{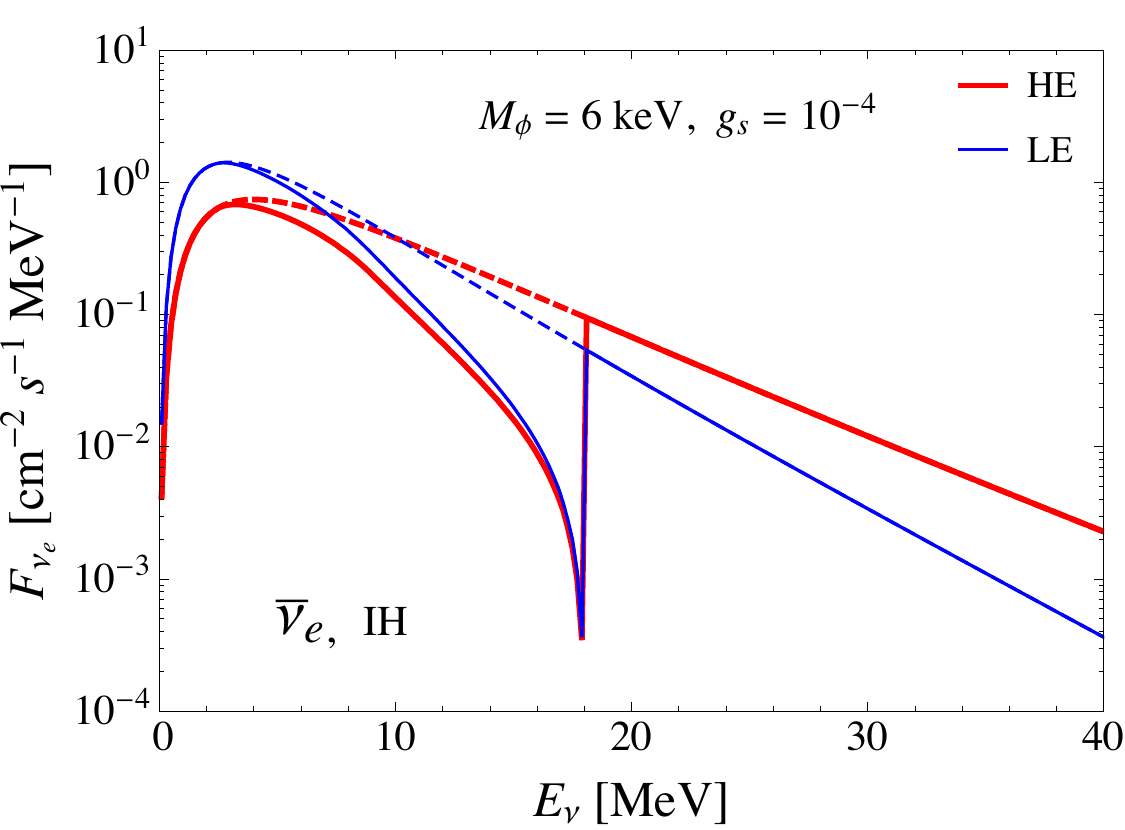}
	\caption{The $\nu_e$ (upper panels) and $\bar{\nu}_e$ (lower panels) DSNB fluxes with (solid curves) and without (dashed curves) resonant interactions, for $M_\phi = 6$~keV, $g_s = 10^{-4}$, $m_s = 1$~eV and $\theta_0 = 0.1$. We show the fluxes for HE (red curves) and LE (blue curves) initial SN neutrino energy spectra and for NH (left panels) and IH (right panels).}
	\label{fig:fnue-int}
\end{figure}

To evaluate the prospects of detecting this type of signal, we consider argon and water-\v{C}erenkov detectors. A liquid argon detector like DUNE would be uniquely sensitive to the $\nu_e$ component, primarily via the charged-current interaction 
\begin{equation}
\nu_e + ^{40} {\rm Ar} \to e^- + ^{40} {\rm K}^*  ~,
\end{equation}
which will provide complementary information to what can be obtained from current or planned water-\v{C}erenkov detectors, which are sensitive primarily to the $\bar\nu_e$ component of the DSNB flux via inverse beta decay, $\bar\nu_e + p \to e^+ + n$. Although the DUNE detector might be sensitive to energies as low as 5~MeV~\cite{Acciarri:2015uup, Ankowski:2016lab}, the main source of background at these low energies is the well-known flux of solar neutrinos. Indeed, the solar \textit{hep} neutrino flux have an endpoint of 18.8~MeV, which implies that distinguishing the DSNB flux at energies below $\sim 16$~MeV would be very challenging~\cite{Cocco:2004ac}. This is similar to the low-energy threshold considered in some Super-Kamiokande analyses~\cite{Malek:2002ns, Bays:2011si}, although spallation backgrounds have been the limiting factor in that case.\footnote{Tagging neutrons in delayed coincidence has been recently used to suppress those backgrounds and lower the positron energy threshold down to 12~MeV~\cite{Zhang:2013tua}.}

With these considerations in mind and recalling that, whereas the absorption would occur in the energy interval $E_{\rm res}/(1 + z_{\rm max}) < E_\nu < E_{\rm res}$, the maximum attenuation would appear at the resonant energy (at $z = 0$), we consider masses of the resonantly produced $\phi$ so that the affected energy range is $10~{\rm MeV} \lsim E_\nu \lsim 30~{\rm MeV}$, i.e., $5~{\rm keV} \lsim M_\phi \lsim 8~{\rm keV}$, for $m_s \simeq 1$~eV. 

In Fig.~\ref{fig:fnue-int}, we depict the attenuated $\nu_e$ (top panels) and $\bar{\nu}_e$ (bottom panels) fluxes. We show spectra for the HE (red curves) and LE (blue curves) initial SN neutrino spectra and for NH (left panels) and IH (right panels). For illustration, all these results are obtained for $M_\phi = 6$~keV, where we can clearly see the strong suppression of the fluxes with respect to the case without attenuation (dashed curves), which extends to energies much below $E_{\rm res} = 18$~MeV. 

\begin{figure}[t]
	\centering
	\includegraphics[width=0.49\textwidth]{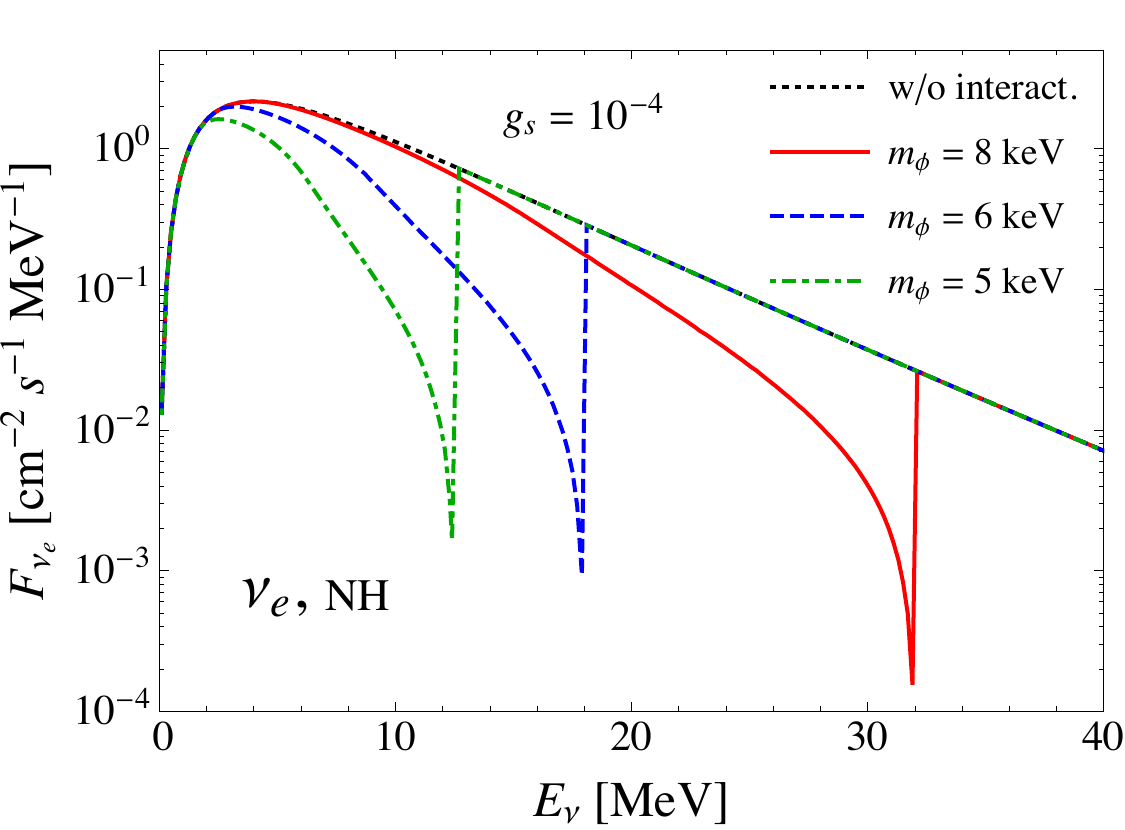}
	\includegraphics[width=0.49\textwidth]{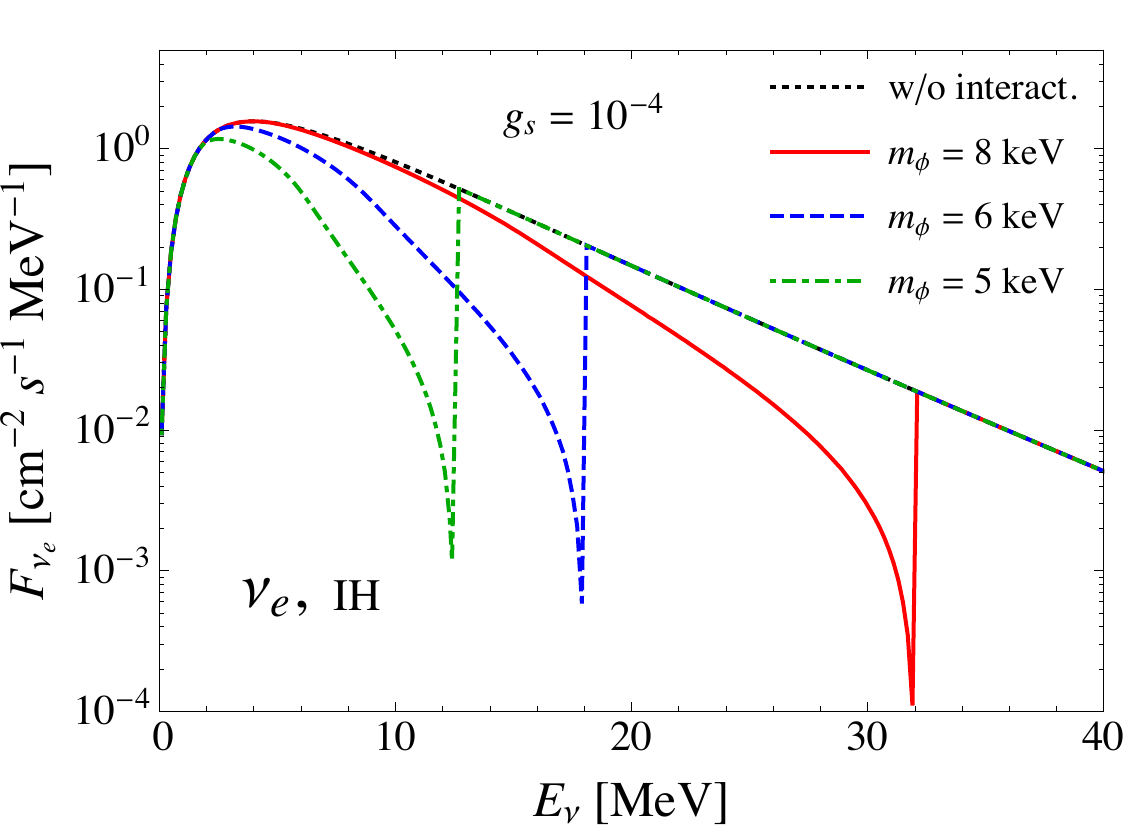}
	\includegraphics[width=0.49\textwidth]{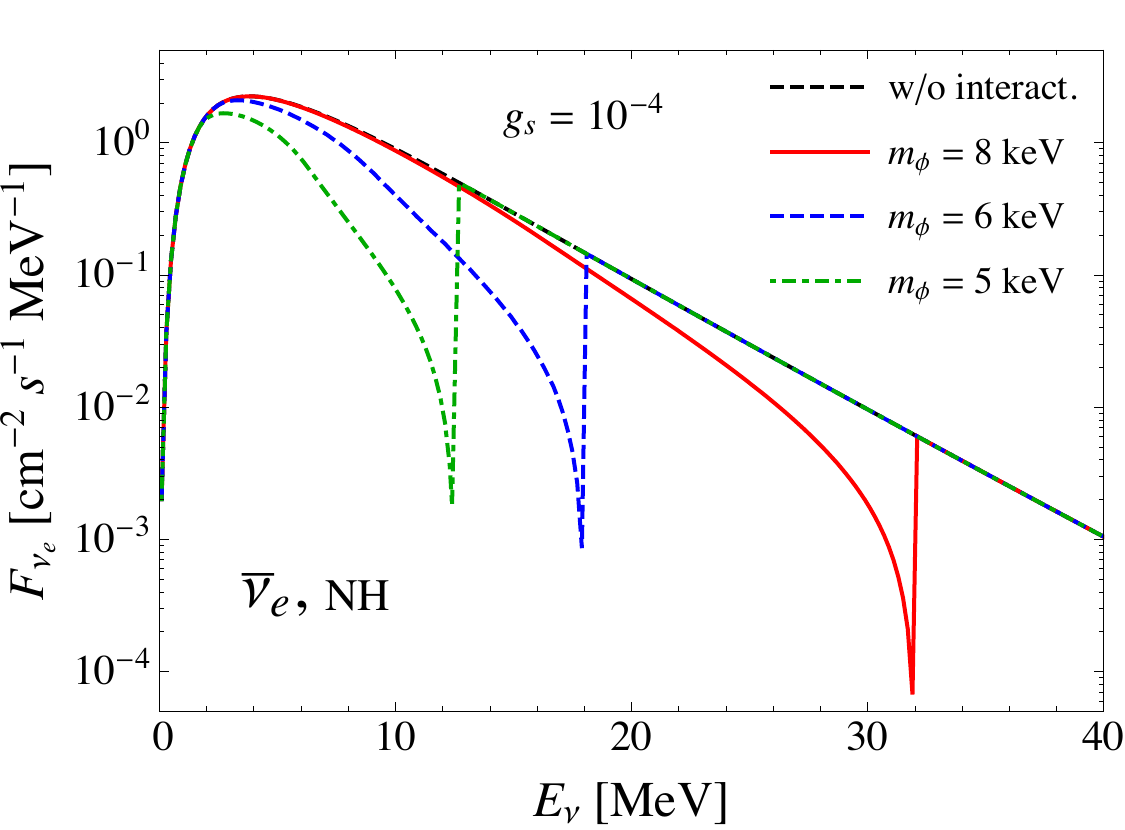}
	\includegraphics[width=0.49\textwidth]{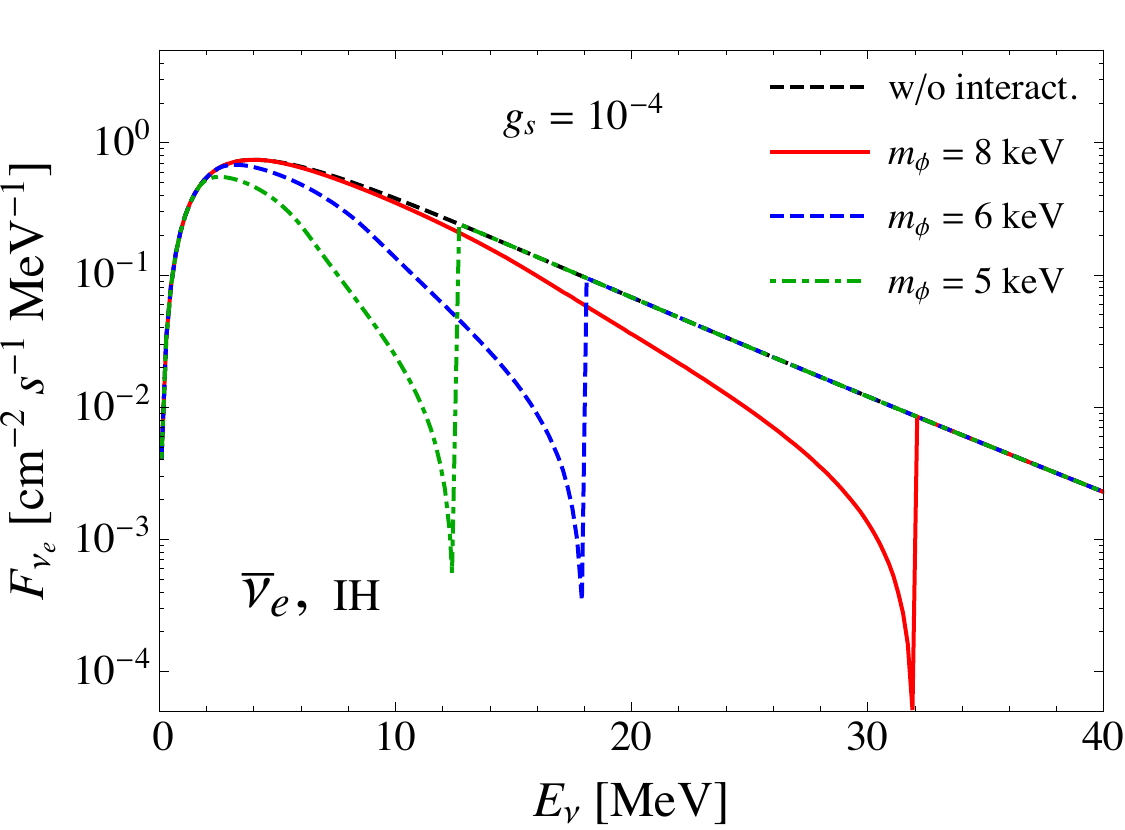}	
	\caption{The $\nu_e$ (upper panels) and $\bar\nu_e$ (lower panels) DSNB flux with and without (black dotted curves) the resonance interactions for the different mediator masses, $M_\phi =$ 5 (green dot-dashed curve), 6 (blue dashed curve) and 8~keV (red solid curve). The spectra are obtained for NH (left panels) and IH (right panels), for the default values $g_s = 10^{-4}$, $m_s = 1$~eV and $\theta_0 = 0.1$.
	}
	\label{fig:fnue-mphi}
\end{figure}

On the other hand, in Fig.~\ref{fig:fnue-mphi} we show the $\nu_e$ (top panels) and $\bar{\nu}_e$ (bottom panels) spectra for different $\phi$ mediator masses ($M_\phi = 5, 6, 8$~keV), for NH (left panels) and IH (right panels), but only for the most optimistic case of initial SN neutrino spectra, i.e., the one with highest average energies (HE).  Even if the flux at the peak in the case of the LE spectra is higher, at energies above the thresholds for detection, the HE fluxes are the most optimistic ones. We can also clearly see how the dip in the SN spectra moves towards higher energies for heavier mediators, where the flux is lower.

\begin{figure}[t]
	\centering
	\includegraphics[width=0.49\textwidth]{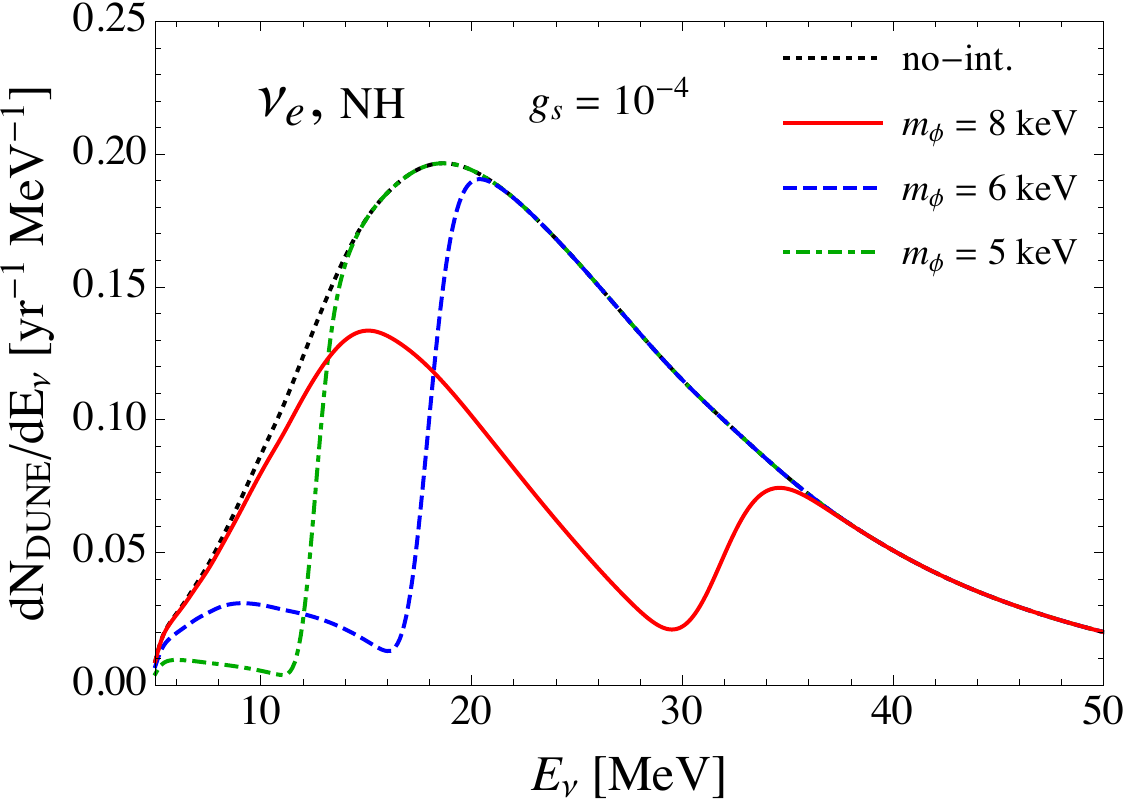}
	\includegraphics[width=0.49\textwidth]{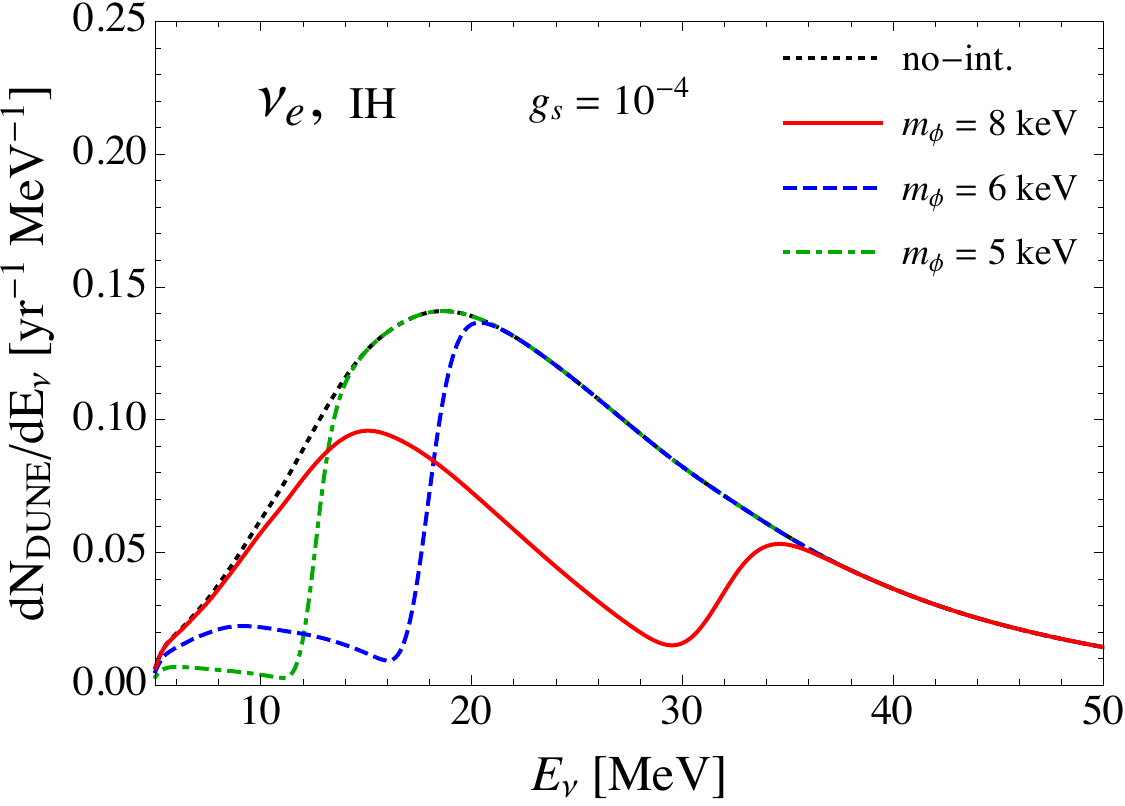}	
	\includegraphics[width=0.49\textwidth]{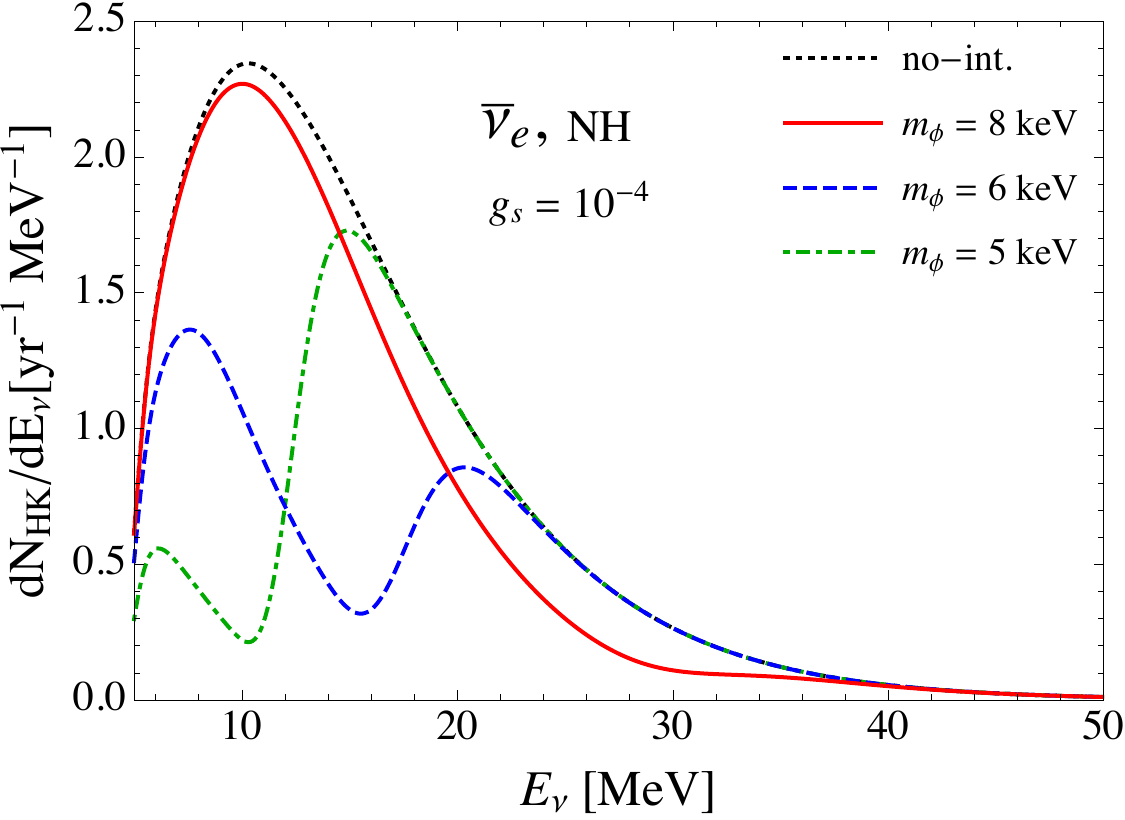}
	\includegraphics[width=0.49\textwidth]{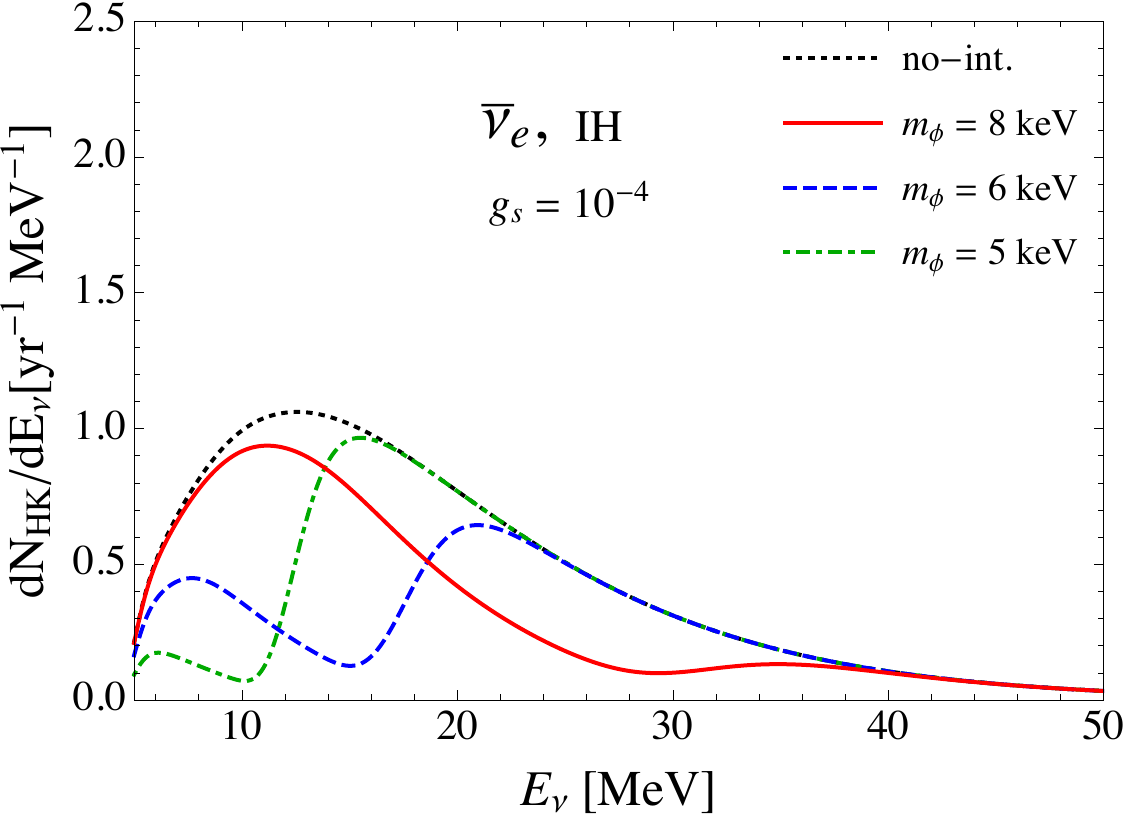}
	\caption{Upper panels: the differential event rates expected at the 40~kton DUNE LAr detector from $\nu_e$ charged-current interactions off $^{40}$Ar, for the same cases depicted in Fig.~\ref{fig:fnue-mphi}. Lower panels: the differential event rates for one 187~kton HK tank from $\bar \nu_e$ inverse beta decay off water.
	}
	\label{fig:Fsigdune}
\end{figure}

Next, we estimate the effects of the new interaction on the shape of the DSNB event spectra in future detectors as DUNE and HK. For the sake of simplicity, we compute the differential event rate as
\begin{equation}
\label{eq:dndeFD2}
\frac{dN_a}{dE_\nu}  = N_T \, \int dE_\nu' \, R(E_\nu, E_\nu') \, F_a (E_\nu') \, \sigma_a (E_\nu') ~,
\end{equation}
where $R(E_\nu, E_\nu')$ is the energy resolution function of the detector and $\sigma_a$ is the reaction cross section corresponding to each detector technique. The total number of targets is given by $N_T$ and we optimistically assume perfect efficiency.

Although a liquid argon detector is sensitive to neutrinos of all three flavors, via charged ($\nu_e$ and $\bar\nu_e$) or neutral current (all flavors) interactions off argon nuclei, or elastic scatterings with atomic electrons (all flavors), the dominant process is $\nu_e + ^{40} {\rm Ar} \to e^-  + ^{40}K^*$. The cross sections for other process are smaller by at least an order of magnitude, so given the low event rate, it would be very challenging to actually detect those other signals from the DSNB flux.

In this work we use the cross section for $\nu_e + ^{40} {\rm Ar} \to e^-  + ^{40}K^*$ computed in the random phase approximation including several multipoles~\cite{Kolbe:2003ys} and that had already been used in earlier works~\cite{Bueno:2003ei, GilBotella:2003sz}. For energies below $\sim 20$~MeV, Fermi and Gamow-Teller transitions are the main components of the cross section~\cite{Ormand:1994js}, whereas other multipoles have to be included at higher energies~\cite{Kolbe:2003ys}. The cross section we use, in agreement with empirical results~\cite{Bhattacharya:2009zz}, is up to a factor of two smaller than other calculations in the literature~\cite{SajjadAthar:2004yf, Suzuki:2012ds, Chauhan:2017tgf}, but it is larger than the calculation of Refs.~\cite{Cheoun:2011zza, Cheoun:2012ha} by a factor of a few. Therefore, more theoretical studies and data seem to be necessary to solve this challenging problem.

For the case of the 40~kton DUNE liquid argon detector ($N_{\rm Ar} = 6 \times 10^{32}$), we consider a Gaussian energy resolution function of width $\sigma/E_\nu = 0.05$, which is a rough approximation of the electron energy resolution in ICARUS~\cite{Amoruso:2003sw}. This is probably an optimistic situation, which assumes the neutrino energy can be reconstructed by measuring the de-excitation gammas and nucleons and the outgoing electron~\cite{Gardiner}. 

In the top panels of Fig.~\ref{fig:Fsigdune}, we present the differential event rates, $dN_{\rm DUNE}/dE_\nu$, as a function of the neutrino energy, for the same cases as in the top panels of  Fig.~\ref{fig:fnue-mphi}. Although the flux decreases as the energy increases for the whole energy range presented in these plots, the cross section rapidly increases. Specifically, the latter grows by about two orders of magnitude in the range $E_\nu = 5 - 20$~MeV and by an order of magnitude from 20~MeV to 50~MeV. Therefore, the combination of the decreasing DSNB flux and the increasing cross section gives rise to event spectra with a maximum at around $E_\nu \sim 20$~MeV (for the HE initial SN neutrino spectrum). The attenuation in the flux caused by the new interactions in the sterile sector would produce a very sharp suppression of the event spectra around the resonance energy. If this is not very close to, and not too far from, the maximum in the spectrum of events without hidden interactions, i.e., if $M_\phi \sim (8-9)$~keV, two peaks would develop, one around the resonance energy and another one just below the maximum in the case of no attenuation, i.e., $E_\nu \sim (10-20)$~MeV. The double peak structure would probably be difficult to mimic by other mechanisms. However, unless the energy threshold is lowered, it would be hard to actually measure both peaks.

In Tables~\ref{table:events16} and~\ref{table:events10}, we indicate the number of events expected after ten years in the DUNE detector for different cases within a model with sterile neutrinos that mix with active ones\footnote{Also notice that similar numbers of events are expected for the standard three-flavor scenario without sterile neutrinos, with a bit larger differences in the case of the IH.} and may have interactions in the sterile sector mediated by a vector boson with mass in the keV range. Although the rate of detection is only of a few events per year, hidden interactions of sterile neutrinos could produce significant features in the observed spectra, with a suppression of up to $50\%$ for $M_\phi = 8$~keV for energies below $\sim 40$~MeV. For smaller masses, the suppression is expected at lower energies and thus, if above threshold, the dip is only expected in the first few bins.

\begin{table}[t]
	\vskip 0.35in
	\begin{center}
		\begin{tabular}{|c|c|c|c|c|c|c|}
			\hline
			DUNE ($\nu_e$)& w/o interaction & $M_\phi$  = 5 keV & $M_\phi$ = 6 keV & $M_\phi$ = 8 keV  & w/o $\nu_s$ \\ 
			\hline
			NH & 32  & 32 & 28 & 16 & 32\\
			\hline
			IH & 23  & 23 & 20 & 12 & 25\\
			\hline
			\hline
			HK ($\bar{\nu}_e$)   & w/o interaction & $M_\phi$  = 5 keV & $M_\phi$ = 6 keV & $M_\phi$ = 8 keV &  w/o $\nu_s$\\ 
			\hline
			NH & 179  & 179 & 133 & 121 & 316\\
			\hline
			IH & 149  & 148 & 120 & 77 & 462\\
			\hline
		\end{tabular}
	\end{center}
	\caption{Number of events expected after the first 10 years of operation of DUNE, $400$~kton$\cdot$yr (top table) and of HK, $2.618$~Mton$\cdot$yr (bottom table), in the energy range $16 {\ \rm MeV} \leq E_\nu \leq 40 {\ \rm MeV}$, for our default SNR~\cite{Yuksel:2008cu, Horiuchi:2011zz, Kistler:2013jza} and for the HE initial neutrino SN spectra~\cite{Lunardini:2010ab}. We show the resulting numbers for the case without new interactions (third column), with new interactions for different values of $M_\phi$ (fourth to sixth columns) and for the standard three-neutrino scenario, using the corresponding flavor-mass eigenstates flux relations~\cite{Dighe:1999bi} (last column).}
	\label{table:events16}
\end{table}

\begin{table}[t]
	\vskip 0.35in
	\begin{center}
		\begin{tabular}{|c|c|c|c|c|c|}
			\hline
			DUNE ($\nu_e$) &   w/o interaction & $M_\phi$  = 5 keV & $M_\phi$ = 6 keV & $M_\phi$ = 8 keV & w/o $\nu_s$ \\ 
			\hline
			NH  & 32  & 29 & 21 & 17 & 32\\
			\hline
			IH & 23  & 21 & 15 & 12  & 27\\
			\hline
			\hline
			HK ($\bar{\nu}_e$) &   w/o interaction & $M_\phi$  = 5 keV & $M_\phi$ = 6 keV & $M_\phi$ = 8 keV & w/o $\nu_s$ \\ 
			\hline
			NH & 337  & 252 & 164 & 273 & 528  \\
			\hline
			IH & 209 & 170 & 111 & 133 &  642 \\
			\hline
		\end{tabular}
	\end{center}
	\caption{Same as Tab.~\ref{table:events16}, but for the energy range $10 {\ \rm MeV} \leq E_\nu \leq 30 {\ \rm MeV}$.}
	\label{table:events10}
\end{table}

Let us now consider the expected DSNB signal in the proposed HK detector set up, with two tanks with a fiducial volume of 187~kton each ($N_{\rm HK} = 1.25 \times 10^{34}$ free protons)~\cite{HK}, but with the second tank only becoming operational after six years of the first one. We will assume a flat energy resolution of 10\% over the entire neutrino energy range, which is similar to the SK energy resolution~\cite{Malek:2002ns, Bays:2011si, Hosaka:2005um, Cravens:2008aa, Abe:2010hy}. Although there is also a small contribution from $\nu_e$ and $\bar\nu_e$ interactions off oxygen nuclei, we will only consider the signal rate from inverse beta decay events off free protons, $\bar\nu_e + p \to e^+ + n$, which is the only relevant one at these energies. We take the total cross section for this process from Tab.~1 in Ref.~\cite{Strumia:2003zx}.

In the bottom panels of Fig.~\ref{fig:Fsigdune}, we show the expected differential event rate per tank, $dN_{\rm HK}/dE_\nu$, as a function of the measured neutrino energy for the same cases considered in the bottom panels of Fig.~\ref{fig:fnue-mphi}. In contrast to the LAr case, the rise of the inverse beta decay cross section is more gradual, so the maximum of the event spectrum (for the HE initial neutrino SN spectrum, without new interactions) in water-\v{C}erenkov detectors is expected at lower energies, $E_\nu \sim 10$~MeV and another maximum could show up at higher energies. On the other hand, the attenuation effect due to the new interaction is expected to be less pronounced than in LAr detectors. However, if there is no mediator or $M_\phi \lsim 5$~keV, the peak (at $\sim 10$~MeV or at the resonant energy) would lie below the energy threshold of $\sim 16$~MeV, so only a falling event spectrum (identical to the case with no resonant interaction) could be detected, unless the threshold is lowered. 

As for the case of DUNE, in Tables~\ref{table:events16} and~\ref{table:events10}, we also indicate the number of events expected in ten years after the first HK tank starts operating (i.e., for an exposure of $2.618$~Mton$\cdot$yr) for different cases and energy intervals. Given the larger size of HK, the expected event rate is of the order of few tens per year. In this case, the suppression could be as much as $30\%$ with respect to the case of no hidden sterile neutrino self-interactions\footnote{However, notice that the differences between the $3+1$ scenario without self interactions and the standard three-neutrino case are much larger than for the DSNB detection in DUNE.}, for $M_\phi = 8$~keV and neutrino energies below $\sim 40$~MeV. We stress that the shape of the event spectra is expected to be different from that at the DUNE detector, which would help to identify this scenario as the cause of the attenuation features. Moreover, for water-\v{C}erenkov detectors, the mass of the vector boson for the most optimistic case depends more importantly on the energy threshold of the experiment.

\section{Summary}
\label{sec:conclusions}

Sterile neutrinos with masses in the eV scale and with large mixings with active neutrinos ($\theta_0 \sim 0.1$) are motivated by anomalies found in short-baseline neutrino experiments~\cite{Athanassopoulos:1995iw, Athanassopoulos:1996jb, Athanassopoulos:1996wc, Aguilar:2001ty, AguilarArevalo:2007it, AguilarArevalo:2010wv, Aguilar-Arevalo:2013pmq, Mention:2011rk}. However, with such large mixings they would be fully thermalized at the BBN time, at odds with current data~\cite{Patrignani:2016xqp}. In this work we have revisited a sterile neutrino portal scenario in which eV-scale sterile neutrinos have self-interactions via a new vector vector boson $\phi$. This new interaction term would induce an effective potential which would suppress oscillations in the early Universe, preventing the equilibration of sterile neutrinos and thus, allowing this scenario to be consistent with bounds from $N_{\rm eff}$ at BBN and CMB~\cite{Hannestad:2013ana, Dasgupta:2013zpn, Archidiacono:2014nda, Saviano:2014esa, Mirizzi:2014ama, Cherry:2014xra, Tang:2014yla, Forastieri:2015paa, Chu:2015ipa, Archidiacono:2015oma, Cherry:2016jol, Archidiacono:2016kkh}. 

In Section~\ref{sec:lagrangian}, within a $V-A$ self-interacting model, we described in detail the ingredients entering the production rate of sterile neutrinos in the early Universe, including all relevant cross sections (see Fig.~\ref{fig:sigs-avg}), the induced effective potential (see Appendix~\ref{app:Veff}) and the quantum damping term in the conversion probability. Note that similar scenarios have been considered in previous works~\cite{Hannestad:2013ana, Dasgupta:2013zpn, Archidiacono:2014nda, Mirizzi:2014ama, Cherry:2014xra, Chu:2015ipa, Cherry:2016jol}, although not all the relevant terms have always been taken into account. In Section~\ref{sec:constraints}, including the effect of the damping term and of resonant neutrino mixing, we showed how different cosmological observations can constrain this model, in terms of the mass of the new vector boson, $M_\phi$, and the self-coupling of sterile neutrinos, $g_s$. In particular, we considered the limits from the inferred number of relativistic degrees of freedom at BBN and from the condition that most active neutrinos must free stream at the time of recombination. The resulting allowed region in the $(M_\phi, g_s)$ plane is shown in Fig.~\ref{fig:constraints}. 

Finally, in Section~\ref{sec:DSNB}, we studied the possibility of detecting the distortion of the DSNB flux due to self-interacting eV-scale sterile neutrinos, within the allowed region discussed in the previous section (also highlighted in Fig.~\ref{fig:constraints}). In order to test these signatures, we considered the expected signals in the planned DUNE~\cite{Acciarri:2015uup, Acciarri:2016crz} and HK detectors~\cite{HK} after ten years of starting taking data. We have first described the predicted DSNB flux by considering different SNR (Fig.~\ref{fig:SNR}) and two different initial SN neutrino spectra (Fig.~\ref{fig:fnue0res-morder} and Table~\ref{table:dndeparam}). We next evaluated the range of $M_\phi$ and $g_s$, allowed by cosmological observations, that would produce a significant dip in the DSNB flux due to the resonant interaction of SN neutrinos off the relic sterile neutrino background. Interactions of diffuse supernova neutrinos with relic sterile neutrinos, can resonantly produce $\phi$'s which then decay into sterile neutrinos, resulting in the depletion of the DSNB flux. In order to potentially detect this distortion of the DSNB flux at future neutrino detectors, the resonant energy must lie in the range from few MeV to tens of MeV, so the mass of the $\phi$ boson would need to be in the range of $\sim 5-10$~keV (for a target mass $m_s = 1$~eV) and in order to have significant dips in the spectrum, the coupling $g_s$ would have to be larger than $10^{-7}$.  To be consistent with the parameter space that is not excluded by cosmology, we considered $M_\phi =$ 5~keV, 6~keV and 8~keV, with a coupling $g_s = 10^{-4}$. The expected attenuation of the DSNB flux is shown in Figs.~\ref{fig:fnue-int} and~\ref{fig:fnue-mphi} for different representative cases. We also note that for larger values of the sterile neutrino mass, the exclusion region in Fig.~\ref{fig:constraints} shifts to larger values of $M_\phi$ and $g_s$, but at the same time, to keep the same resonant neutrino energy, the value of $M_\phi$ must also be larger. Therefore, a vertical band in the parameter space (analogous to the orange region in Fig.~\ref{fig:constraints}, but shifted to higher $M_\phi$ and $g_s$) might still be testable with future DSNB data.

We find the signals at HK and DUNE to be unique for both normal and inverted light neutrino mass hierarchies (Fig.~\ref{fig:Fsigdune}). Nevertheless, our results should be taken as an illustration of the potential signals in these detectors, given the large uncertainties in the predicted fluxes and detection cross sections (in particular for DUNE). On one hand, the expected energy spectra of SN neutrinos would give rise to event spectra that vary within a factor of a few in the relevant energy range. In addition, the SNR is only estimated within a factor of a few. Moreover, there could be a significant fraction of failed SN which could enhance the chances of detection for $E_\nu > 20$~MeV~\cite{Lunardini:2009ya}. Finally, the electron neutrino scattering cross section off argon nuclei relevant for the DUNE detector is only known within a factor of a few~\cite{Ormand:1994js, Kolbe:2003ys, SajjadAthar:2004yf, Bhattacharya:2009zz, Cheoun:2011zza, Cheoun:2012ha, Suzuki:2012ds, Chauhan:2017tgf}. All in all, these uncertainties translate into variations of the predicted event rates by about an order of magnitude. Bearing these caveats in mind, our results indicate that whereas HK could detect several tens of $\bar\nu_e$-induced events per year, the DUNE detector might only detect about a few $\nu_e$-induced events per year, given its smaller size. For the most optimistic of the cases illustrated in this work and in the energy range $E_\nu = (16, 40)$~MeV, the suppression of the event rate with respect to the case without new interactions could be at the level of $\sim 50\%$ and $30\%$ in the DUNE and HK detectors, respectively. We stress that the total number of events could also be larger if the energy threshold in these detectors is reduced with respect to the default one use in this work, $E_\nu \sim 16$~MeV (see Tables~\ref{table:events16} and~\ref{table:events10}).

In conclusion, we have argued that self-interacting eV-scale sterile neutrinos are not only an allowed possibility from cosmological observations, but in some regions of the parameter space $(M_\phi, g_s)$, their existence can be probed by future observations of the DSNB in planned detectors, as that designed for the DUNE experiment or the water-\v{C}erenkov HK detector.

\acknowledgments

We thank Kate Scholberg for discussions about the DUNE detector. This research was supported in part by the US Department of Energy contracts DE-SC-0010113, DE-SC-0010114, DE-SC-0002145, DE-SC0009913. SPR is supported by a Ram\'on y Cajal contract, by the Spanish MINECO under grants FPA2017-84543-P, FPA2014-54459-P and SEV-2014-0398, by the Generalitat Valenciana under grant PROMETEOII/2014/049 and by the European Union's Horizon 2020 research and innovation programme under the Marie Sk\l odowska-Curie grant agreements No 690575 and No 674896. SPR is also partially supported by the Portuguese FCT through the CFTP-FCT Unit 777 (PEst-OE/FIS/UI0777/2013).

\appendix
\section{Effective potential}
\label{app:Veff}

For completeness, we include here the result for the effective potential for sterile neutrino scattering with sterile neutrino background at temperature $T_s$, which has already been computed in Ref.~\cite{Dasgupta:2013zpn}. This is analogous to the SM neutrino effective potential~\cite{Weldon:1982bn, Notzold:1987ik, DOlivo:1992lwg, Quimbay:1995jn, Konstandin:2005vx}, with the $W$ or $Z$ boson replaced by the $\phi$ boson with coupling $g_s$. Here, we provide the details of this calculation and make some further comments about it.

In general, the propagation of sterile neutrinos, $\nu_s$, is governed by the Dirac equation,
\begin{equation}
\left[ \slashed K - \Sigma (K) \right] \psi_s = 0 ~, 
\end{equation}
where $\slashed K = \gamma_\mu K^\mu$, $K^\mu$ is the neutrino four-momentum\footnote{For momentum vectors, we will use capital letters for the four-momentum and lowercase letters for the modulus of the three-momentum, e.g., $k \equiv |\vec{k}|$.} and $\Sigma (K)$ is the $\nu_s$ self-energy. In vacuum, $\Sigma (K)$ is proportional to $K$ and thus, the pole of the propagator is located at $K^2 = m_s^2$, where $m_s$ is the (mostly) sterile neutrino mass. However, in a bath of particles at finite temperature, $T_s$, there is a preferential frame, the center-of-mass frame of the plasma. In general, this frame has four-velocity $u^\mu$, with $u^\mu u_\mu = 1$. Therefore, the self-energy for left-handed fermions in the presence of a medium is of the general form:\footnote{At one loop, terms proportional to $\slashed K \slashed u$ are not generated~\cite{Quimbay:1995jn}.}
\begin{equation}
\Sigma(k) = m_s - a \slashed K \, P_L - b \slashed u \, P_L ~, 
\end{equation}
where $P_L = (1 - \gamma_5)/2$ and $a$ and $b$ are functions of the Lorentz-invariant quantities,
\begin{eqnarray}
\omega & = & K^\mu u_\mu ~, \nonumber \\
k & = & \left( w^2 - K^2 \right)^{1/2} ~.
\end{eqnarray}
It is convenient to split the self-energy into zero-temperature and finite-temperature contributions, $\Sigma (K) = \Sigma_0 (K) + \Sigma_T (K)$. The $\Sigma_0 (K)$ part does not contribute to the dispersion relation and it only renormalizes the wave function, so from now on, we will only consider the background-dependent part, $\Sigma_T (K)$. In general, $\Sigma_T (K)$ is a complex quantity, but we are only interested in its real part.\footnote{The imaginary part, being of order $k$, is related to the damping rate of the particle in the thermal bath~\cite{Braaten:1992gd}.}

The poles of the fermion propagator determine the dispersion relation, which is equivalent to solving $\rm{det}(\slashed K - \Sigma (K)) = 0$, so that non-trivial solutions of the Dirac equation exist. For relativistic neutrinos, i.e., $k \gg m$, and up to first order in small quantities, 
\begin{equation}
w \simeq k +\frac{m_s^2}{2 \, k} - b \simeq  k +\frac{m_s^2}{2 \, k} + V_{\rm eff} (k, T_s) ~,
\end{equation}
so $V_{\rm eff} (k, T_s) \simeq - b(w, k, T_s)$ can be treated as the effective potential induced by the presence of the medium, which can be expressed as
\begin{equation}
V_{\rm eff} = \frac{1}{2 \, k^2} \, \left[ \omega \,  {\rm Tr}\{\slashed K \, Re \Sigma_T(K)\} - (\omega^2 - k^2) \, {\rm Tr}\{\slashed u \, Re \Sigma_T(K)\}\right] ~. 
\label{eq:b}
\end{equation}

Considering an interaction term as that in Eq.~(\ref{eq:Ls}), i.e., ${\cal L}_{s} = g_s \, \bar{\nu}_s \gamma_\mu  P_L \, \nu_s \, \phi^\mu$, between $\nu_s$ and a vector boson $\phi$, with mass $M_\phi$, and assuming the same particle distributions for neutrinos and antineutrinos, the effective potential comes only from the bubble diagram of the $\nu_s$ self-energy. The tadpole diagram only contributes in case of an asymmetry. The bubble diagram for the $\nu_s$ self-energy is given by
\begin{equation}
\Sigma (K) = -i g_s^2 \int \frac{d^4P}{(2 \, \pi)^4} \, \gamma^\mu \, P_L \, i \, S(P + K) \, \gamma^\nu \, P_L \, i \, D_{\mu \nu} (P) ~,
\end{equation} 
where $g_s$ is the $\nu_s-\phi$ coupling and the thermal propagators for massive fermions and vector bosons (in the unitary gauge) are given by
\begin{eqnarray}
S(P) & = & (\slashed p + m_s) \left[ \frac{1}{P^2 - m_s^2 + i \epsilon} + i \, \Gamma_f (P)\right] ~, \\
D_{\mu \nu} (P) & = & \left(- g_{\mu \nu} + \frac{P_\mu \, P_\nu}{M_\phi^2} \right) \, \left[\frac{1}{P^2 - M_\phi^2 + i \epsilon} - i \, \Gamma_b(P) \right] ~.
\end{eqnarray} 
The temperature dependence occurs via the functions $\Gamma_f (p)$ and $\Gamma_b (p)$, defined as
\begin{equation}
\Gamma_\beta (P) = 2 \, \pi \, \delta(P^2  - M_\phi^2) \, f_\beta(P)
\end{equation}
with 
\begin{equation}
f_\beta (P) = \frac{1}{e^{(|P \cdot u| - \sgn(P \cdot u) \, \mu)/ T_s} - \beta} ~, 
\end{equation}
where $\beta = 1$ and $\beta = -1$ correspond to bosons and fermions, respectively. Given that we  assume the particle and antiparticle distributions to be equal, $\mu = 0$. When convenient, we will use the subindexes $b$ and $f$ instead of $+$ and $-$ (for $\beta = 1$ and $\beta = -1$, respectively).

Following Eq. (\ref{eq:b}),, after some algebra, the effective potential can be written as
\begin{eqnarray}
V_{\rm eff} (k, T_s) & = & g_s^2 \int \frac{dp}{16 \, \pi^2} \, \frac{p}{k^3} \sum_{\beta=1,-1} \frac{1}{E_\beta} \, \left[ 8 \, p \, k \, \omega - 2 \, (\omega^2 - k^2) \, E_\beta \, L_\beta^-(p) \right. \nonumber \\
& & \left. - \, \frac{\omega}{2} \, \left((\omega^2 - k^2 - m_s^2) \, (2 - \beta) + 2 \, \beta \, \Delta + 2 \, m_s^2 \right) \, L_\beta^+(p) \right. \nonumber \\
& & \left. - \, \frac{1}{M_\phi^2} \, \Big\{ 4 \, p \, k \, \omega \left( \omega^2 - k^2 - m_s^2 \right) - \, \omega \, \frac{(\omega^2 - k^2 - m_s^2) \, (\omega^2 - k^2 - \beta \, m_s^2)}{2} \, L_\beta^+(p) \right. \nonumber \\
& & \left. - \, \beta \, E_\beta \, (\omega^2 - k^2 - m_s^2) \, (\omega^2 - k^2) \, L_\beta^-(p) \Big\} \right] \, f_\beta(p) ~,
\label{eq:Veff}
\end{eqnarray}
where $\Delta \equiv M_\phi^2 - m_s^2 = m_b^2 - m_f^2$, $E_\beta\equiv (p^2+m_\beta^2)^{1/2}$ and 
\begin{equation}
L_{\beta}^{\pm}(p) \equiv \ln\left[ \frac{\omega^2 - k^2 + \beta \, \Delta + 2 \, E_\beta \, \omega + 2 \, k \, p}{\omega^2 - k^2 + \beta \, \Delta + 2 \, E_\beta \, \omega - 2 \, k \, p}\right] 
\pm  \ln\left[ \frac{\omega^2 - k^2 + \beta \, \Delta - 2 \, E_\beta \, \omega + 2 \, k \, p}{\omega^2 - k^2 + \beta \, \Delta - 2 \, E_\beta \, \omega - 2 \, k \, p} \right] ~.
\end{equation}

In the high temperature limit, the logarithms go as
\begin{eqnarray}
L_\beta^+(p) & \simeq & - \frac{2 \, k}{p} \, \left( 1 + \frac{\beta \, \Delta}{\omega^2 - k^2}\right) ~, \nonumber \\
L_\beta^-(p) & \simeq & 2 \, \ln\left(\frac{\omega + k}{\omega - k}\right) ~,
\end{eqnarray}
and hence, for $T_s \gg M_\phi, k$,
\begin{eqnarray}
V_{\rm eff} (k, T_s) & \simeq & g_s^2 \int \frac{dp}{2 \, \pi^2} \, \left(\frac{p}{k}\right) \, \Bigg\{   \\
& & + \,  \left(1 - \frac{1}{2} \, \left(\frac{\omega^2}{k^2} -1 \right) \, \ln\left(\frac{\omega + k}{\omega - k}\right) - \frac{\omega^2 - k^2 - m_s^2}{2 \, M_\phi^2}  \right) \, \Big(f_f(p) + f_b(p) \Big) \nonumber \\
& & - \, \frac{(\omega^2 - k^2 - m_s^2)}{4 \, M_\phi^2} \, \left(\frac{\omega^2}{k^2} - 1 \right) \, \ln\left(\frac{\omega + k}{\omega - k}\right) \, \Big(f_f(p) - f_b(p) \Big) \Bigg\} \nonumber \\
& = & \frac{g_s^2 \, T_s^2}{8\, k} \, \left[ 1 - \frac{1}{2} \, \left(\frac{\omega}{k^2} -1 \right) \, \ln\left(\frac{\omega + k}{\omega - k}\right) \right. \nonumber  \\
& & \left. - \, \frac{\omega^2 - k^2 - m_s^2}{2 \, M_\phi^2} + \frac{(\omega^2 - k^2 - m_s^2)}{12 \, M_\phi^2} \, \left(\frac{\omega^2}{k^2} - 1 \right) \, \ln\left(\frac{\omega + k}{\omega - k}\right) \right] ~. \nonumber 
\end{eqnarray}
Thus, for high-energy neutrinos ($\omega \simeq k$), up to first order, the potential is given by
\begin{equation}
V_{\rm eff} (k, T_s) \simeq \frac{g_s^2 \, T_s^2}{8 \, k} ~,
\end{equation}
which agrees with the result in Ref.~\cite{Dasgupta:2013zpn}.

In the limit of a very massive vector boson, $M \gg T_s \gg m_s$, we have
\begin{eqnarray}
L_\beta^+(p) & \simeq & \frac{8 \, k \, p}{M_\phi^2} \, \left( \beta + \beta \, \frac{4}{3} \, \frac{k^2 \, p^2}{M_\phi^4} \, \left(1 + 3 \, \frac{\omega^2}{k^2}\right) - \frac{\omega^2 - k^2}{M_\phi^2} \right) + {\cal O}\left(\frac{k^5 \, p^5}{M_\phi^{10}}, \frac{k^3 \, p^3 \, (\omega^2 - k^2)}{M_\phi^{8}}\right)
~, \nonumber \\
L_\beta^-(p) & \simeq & - \frac{16 \, k^2 \, p^2}{M_\phi^4} \, \frac{\omega}{k} + {\cal O}\left(\frac{k^4 \, p^4}{M_\phi^{8}}, \frac{k^2 \, p^2 \, (\omega^2 - k^2)}{M_\phi^{6}}\right) ~.
\end{eqnarray}
After cancellations at the lowest order and noting that in this limit the number vector bosons in the bath is Boltzmann suppressed, the dominant term in Eq.~(\ref{eq:Veff}) is $-\beta \, \omega \, \Delta \, L_f^+(p)$ (only $\beta = -1$ needs to be considered), so we get
\begin{equation}
V_{\rm eff} \simeq -\frac{g_s^2 \, 2 \, k}{3 \, \pi^2 \, M_\phi^4} \int_0^\infty dp \, p^3 \, f_f(p)  = - \frac{g_s^2 \, 7 \, \pi^2 \, k\, T_s^4}{45 \, M_\phi^4} ~,
\end{equation}
which also agrees with the corresponding result of Ref.~\cite{Dasgupta:2013zpn}.

In order to compute the effective potential in the intermediate regime, one simplification can be made. Assuming the mass of the fermion to be much smaller than any other scale in the problem, we can neglect all $m_s^2$ and $(\omega^2 - k^2)$ (and obviously $(\omega^2 - k^2 - m_s^2)$) terms in Eq.~(\ref{eq:Veff}), so the final expression for the effective potential, which is used throughout this paper, reads
\begin{eqnarray}
V_{\rm eff} (k, T_s) & \simeq & 
g_s^2 \int \frac{dp}{16 \, \pi^2} \, \frac{p}{k^2} \sum_\beta \frac{1}{E_\beta} \, \left[ 8 \, p \, k  - \beta  \, \Delta \, L_\beta^+(p) \right] \, f_\beta(p) \nonumber \\
& \simeq & g_s^2 \int \frac{dp}{16 \, \pi^2} \, \left[ \left( 8 \, \frac{p}{k}  +  \frac{M_\phi^2}{k^2} \, L_f^+(p) \right) \, f_f(p) + \frac{p}{E_b} \, \left( 8 \, \frac{p}{k}  - \frac{M_\phi^2}{k^2} \, L_b^+(p) \right) \, f_b(p) \right]  \nonumber ~, 
\label{eq:Veff2}
\end{eqnarray}
where
\begin{eqnarray}
L_f^{+}(p) & \simeq & \ln\left[ \frac{M_\phi^2 - 4 \, k \, p}{M_\phi^2 + 4 \, k \, p}\right] \nonumber \\
L_b^{+}(p) & \simeq & \ln\left[ \frac{M_\phi^2 + 2 \, k \, (E_b + p)}{M_\phi^2 + 2 \, k \, (E_b - p)}\right] + \ln\left[ \frac{M_\phi^2 - 2 \, k \, (E_b - p)}{M_\phi^2 - 2 \, k \, (E_b + p)} \right] ~,
\end{eqnarray}
with $E_b = (p^2 + M_\phi^2)^{1/2}$. Accounting for the finite width $\Gamma_\phi$ of the $\phi$ requires the substitution in Eq.~(\ref{eq:Veff2}),
\begin{equation}
L_f^{+}(p) \to \frac{1}{2}\ln\left[ \frac{(M_\phi^2 - 4 \, k \, p)^2 + \Gamma_\phi^2 \, M_\phi^2}{(M^2 + 4 \, k \, p)^2 + \Gamma_\phi^2 \, M_\phi^2}\right] 
- \frac{\Gamma_\phi}{M_\phi}\left[ \tan^{-1}\left( \frac{M_\phi^2 + 4 \, k\, p }{\Gamma_\phi \, M_\phi} \right)
- \tan^{-1}\left( \frac{M_\phi^2 - 4 \, k \, p }{\Gamma_\phi \, M_\phi} \right) \right] ~.
\end{equation}
Note that the high-energy and low-energy limits of $V_{\rm eff}$ are unchanged with this substitution.

Finally, note that the effective potential is gauge invariant up to the order we are considering, ${\cal O}(m^2/k^2)$. Recall that we have computed it in the unitary gauge. We could repeat the exercise for the gauge-dependent part of the boson propagator, which in a general gauge can be written as
\begin{equation}
D_{\mu \nu}^{\xi} (P) = - \frac{1}{M_\phi^2} \, \frac{P_\mu \, P_\nu}{P^2 - M_\phi^2/\xi} ~,
\end{equation}
so that we exactly recover our previous result for $\xi = 0$. Notice that the structure of this propagator is analogous to that of the longitudinal part in the unitary gauge and that, up to the order we consider, only the transversal part of the propagator contributes to the effective potential. Thus, up to ${\cal O}(m_s^2/k^2)$, although the residue of the $\xi$-dependent propagator (and thus, the $\nu_s$ self-energy), which is related to the wave-function normalization, depends on the gauge, the pole of the propagator, which is related to the dispersion relation, is gauge independent.

\begin{figure}
	\includegraphics[width=\textwidth]{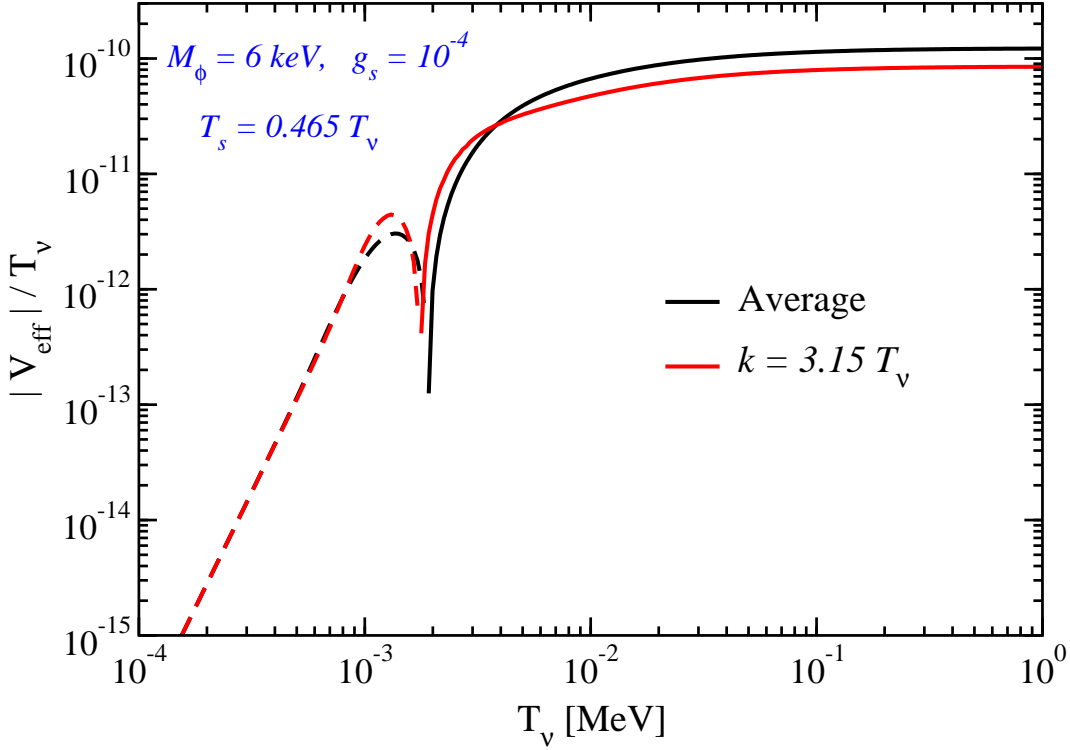} 
	\caption{Effective potential as a function of the temperature of the oscillating neutrino, $T_\nu$, for a sterile neutrino background distribution with temperature $T_s = 0.465 \, T_\nu$: averaging over the neutrino distribution (black curves) and for $k = 3.15 \, T_\nu$ (red curves). Solid curves correspond to $V_{\rm eff} > 0$ and dashed curves to $V_{\rm eff} < 0$. The results are shown for our benchmark values $M_\phi = 6$~keV and $g_s = 10^{-4}$.}
	\label{fig:Veff}
\end{figure}

In Fig.~\ref{fig:Veff} we show the effective potential as a function of the temperature $T_\nu$ of the sterile neutrino state that oscillates with the active one, for a sterile neutrino background distribution with temperature $T_s = 0.465 \, T_\nu$, and for $M_\phi = 6$~keV and $g_s = 10^{-4}$, our benchmark values in this work. The dashed curves correspond to $V_{\rm eff} < 0$, while the solid lines represent $V_{\rm eff} > 0$. The red curves correspond to a momentum fixed to the average value for a thermal distribution with temperature $T_\nu$, i.e., $k = 3.15 \, T_\nu$, while the black curves show the $k$-average over a thermal distribution of neutrinos with temperature $T_\nu$, i.e.,
\begin{equation}
\langle V_{\rm eff} \rangle (T_\nu, T_s) = \frac{\int dk \, k^2 \, V_{\rm eff}(k, T_s) \, f_f(k, T_\nu)}{\int dk \, k^2 \, f_f(k, T_\nu)} ~.
\end{equation}
As can be seen, both results are very similar. Indeed, in the low-temperature limit, they coincide exactly. Given our approximate arguments concerning the observational constraints, to avoid having to compute for every point in the parameter space the thermal average of the different relevant quantities, as the mixing angle in the medium, all the results in this paper are obtained fixing the neutrino momentum at its thermal average value, i.e., $V_{\rm eff}(k, T_s) \to V_{\rm eff} (k = 3.15 \, T_\nu, T_s)$.

\clearpage
\bibliographystyle{JHEP}
\bibliography{refs}

\end{document}